%% file: msxxx061130.tex
\documentclass[]{emulateapj}
\usepackage{amssymb,amsmath}
\usepackage{natbib}

\newcommand{\myemail}{pedro@ifa.hawaii.edu}
\newcommand{\tna} {\,\tablenotemark{a}}
\newcommand{\tnb} {\,\tablenotemark{b}}
\newcommand{\tnc} {\,\tablenotemark{c}}
\newcommand{\tnd} {\,\tablenotemark{d}}
\newcommand{\tne} {\,\tablenotemark{e}}
\newcommand{\tnf} {\,\tablenotemark{f}}
\newcommand{\tng} {\,\tablenotemark{g}}
\newcommand{\tnh} {\,\tablenotemark{h}}
\newcommand{\tni} {\,\tablenotemark{i}}

\newcommand{\kgm} {$\,$kg$\,$m$^{-3}$ }

\shorttitle{Densities of solar system objects}
\shortauthors{Lacerda and Jewitt}

\begin{document}

\def\FigAvsN{
  \begin{figure}
    \centering
    \includegraphics[width=1.0\columnwidth]{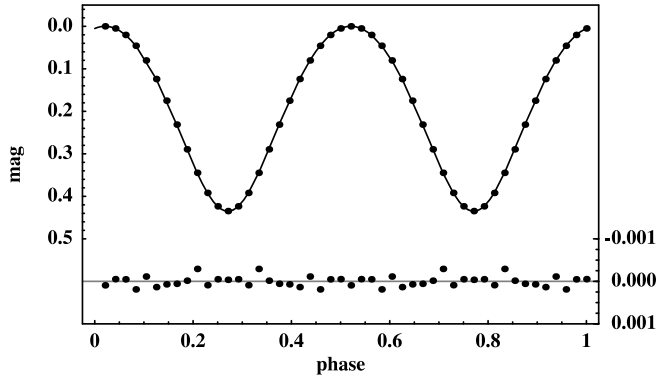}
    \caption[f1.eps] {Accuracy of raytraced versus analytical solution, for an ellipsoid with a ``lunar'' type surface, an axis ratio $b/a$=2/3, observed at aspect angle $\theta=90^\circ$ and phase angle $\alpha=0^\circ$. Bottom points and right ordinate axis show difference between the points (raytraced) and the line (analytical) shown at the top.} 
	\label{Fig.AvsN}
  \end{figure}
}

\def\FigLCurSLawPAngEq{
  \begin{figure}
    \centering
    \includegraphics[width=1.0\columnwidth]{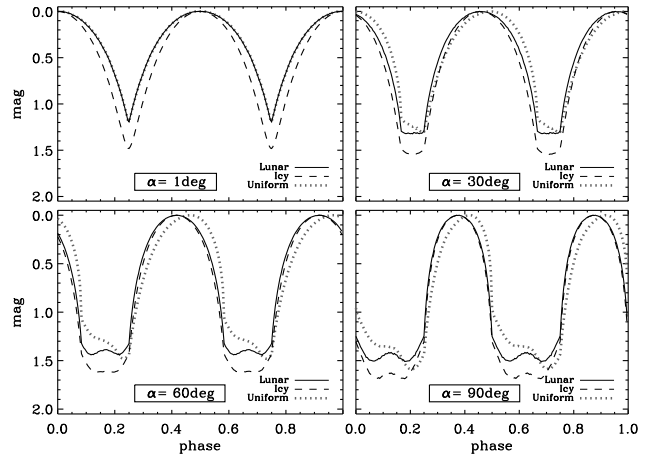}
    \caption[f2.eps] {Lightcurves of a Roche contact binary at different phase angles and for different scattering laws. In the "uniform" case every point of the surface that is illuminated by sunlight reflects exactly the same amount of light back to the observer. Both primary and secondary have axis ratios $b/a=0.67$ and $c/a=0.60$, and the components are in contact.} 
	\label{Fig.LCurSLawPAngEq}
  \end{figure}
}

\def\FigLCurSLawPAngUn{
  \begin{figure}
    \centering
    \includegraphics[width=1.0\columnwidth]{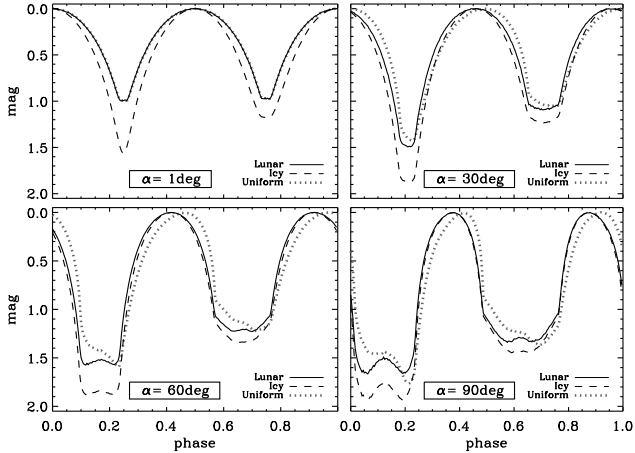}
    \caption[f3.eps] {Same as Fig. \ref{Fig.LCurSLawPAngEq} but for a binary with different size components. Mass ratio is $q=0.67$ and the components have axis ratios ($B/A=0.77,C/A=0.69$) and ($b/a=0.53,c/a=0.49$).} 
	\label{Fig.LCurSLawPAngUn}
  \end{figure}
}

\def\FigLCurSLawPAngAr{
  \begin{figure}
    \centering
    \includegraphics[width=1.0\columnwidth]{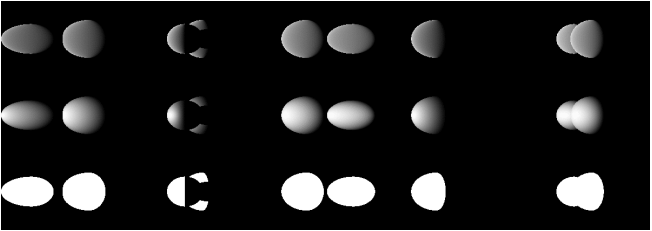}
    \caption[f4.eps] {Three-dimensional rendering of the binary used to produce Fig. \ref{Fig.LCurSLawPAngUn} ($\alpha=60\,$deg). Rotational phase ($\phi=0,72,144,216,$ and $288\,$deg) runs from left to right and top to bottom rows show lunar, icy, and uniform surface scattering functions.} 
	\label{Fig.LCurSLawPAngAr}
  \end{figure}
}

\def\FigHePm{
  \begin{figure}
    \centering
    \includegraphics[width=1.0\columnwidth]{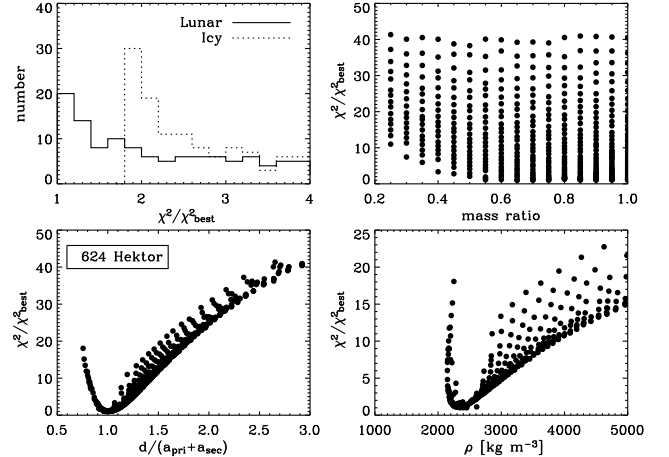}
    \caption[f5.eps] {Quality of fit as a function of scattering function
(top-left), mass ratio of binary components (top-right), orbital distance
(bottom-left), and bulk density of binary components (bottom-right) for (624)
Hektor. To avoid cluttering, only Roche models with mass ratio values, $q$,
multiple of 0.05 are plotted.} 
	\label{Fig.HePm}
  \end{figure}
}

\def\FigHeLC{
  \begin{figure}
    \centering
    \includegraphics[width=1.0\columnwidth]{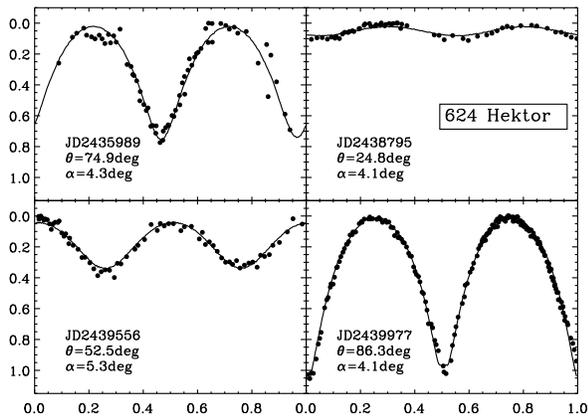}
    \caption[f6.eps] {Roche binary model lightcurve superimposed on data for (624) Hektor taken at 4 different aspects.} 
	\label{Fig.HeLC}
  \end{figure}
}

\def\FigHeSh{
  \begin{figure*}
    \centering
    \includegraphics[width=0.7\textwidth]{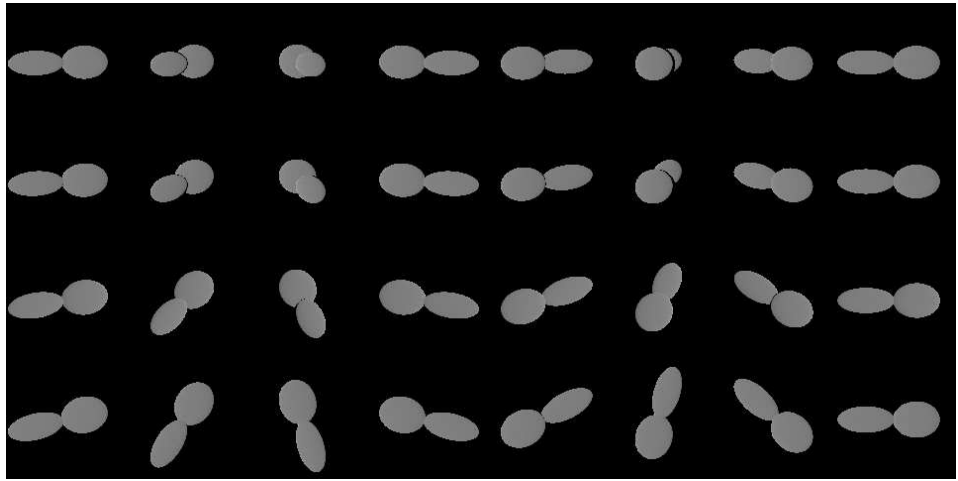}
    \caption[f7.eps] {Rendering of best-fit Roche binary model for (624) Hektor for the 4 geometries listed in Table~\ref{Table.HektorObs}. Rotational phase ($\phi=15, 60, 105, 165, 210, 255, 315,$ and $360\,$deg) runs from left to right while aspect angle decreases from top to bottom.} 
	\label{Fig.HeSh}
  \end{figure*}
}

\def\FigQGPm{
  \begin{figure}
    \centering
    \includegraphics[width=1.0\columnwidth]{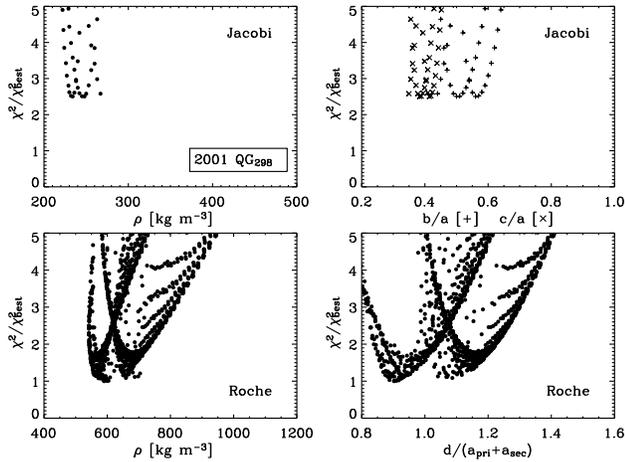}
    \caption[f8.eps] {Quality of fit versus bulk density (top-left) and axis ratios (top-right) of Jacobi ellipsoid models, and versus bulk density (bottom-left) and orbital distance (bottom-right) of Roche binary models for 2001$\,$QG$_{298}$ lightcurve data. Jacobi ellipsoid models are plotted for all four combinations of surface properties and observational geometry listed in Table~\ref{Table.QG}, while Roche binary models are plotted for both lunar- and icy-type surfaces at an aspect angle $\theta=90^\circ$.} 
	\label{Fig.QGPm}
  \end{figure}
}

\def\FigQGLC{
  \begin{figure}
    \centering
    \includegraphics[width=1.0\columnwidth]{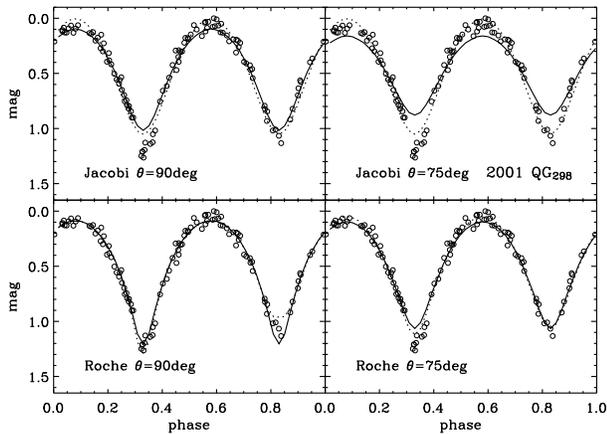}
    \caption[f9.eps] {Models that best fit the lightcurve of 2001$\,$QG$_{298}$ for each combination of scattering law and geometry. Solid and dotted lines indicate lunar and icy surface scattering models, respectively.} 
	\label{Fig.QGLC}
  \end{figure}
}

\def\FigQGSh{
  \begin{figure}
    \centering
    \includegraphics[width=1.0\columnwidth]{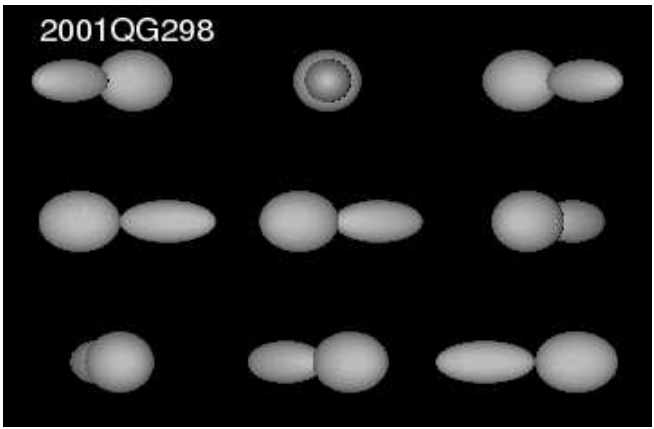}
    \caption[f10.eps] {Rendering of best-fit Roche binary model for 2001$\,$QG$_{298}$. Rotational phase ($\phi=45, 90, 135, 165, 210, 240, 285, 315,$ and $360\,$deg) runs from left to right and top to bottom.} 
	\label{Fig.QGSh}
  \end{figure}
}

\def\FigGNPm{
  \begin{figure}
    \centering
    \includegraphics[width=1.0\columnwidth]{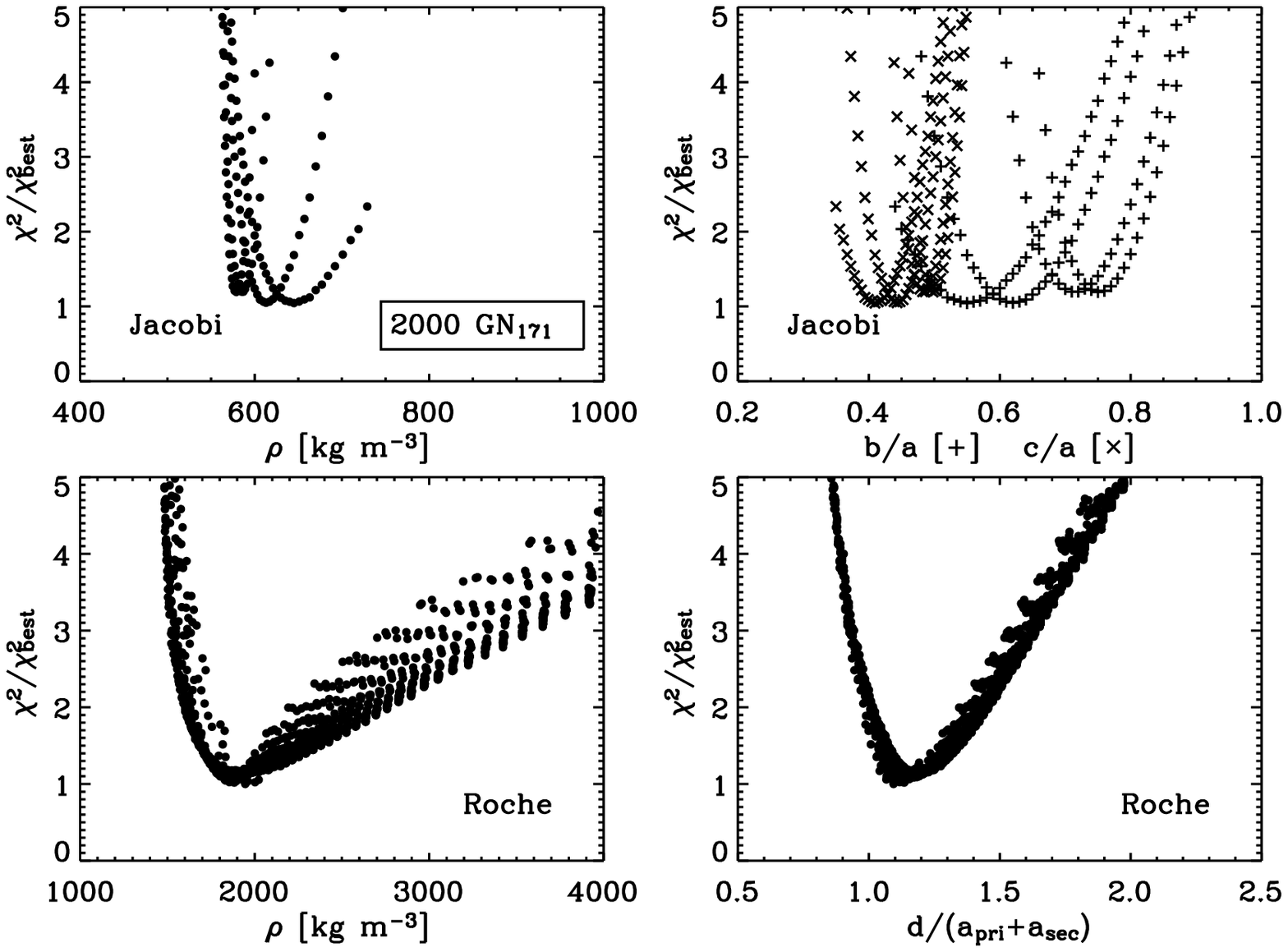}
    \caption[f11.eps] {Same as Fig.~\ref{Fig.QGPm} but for 2000$\,$GN$_{171}$ lightcurve data. Jacobi ellipsoid models are plotted for all four combinations of surface properties and observational geometry listed in Table~\ref{Table.GN}, while Roche binary models are plotted for lunar-type surface and aspect angle $\theta=75^\circ$.} 
	\label{Fig.GNPm} 
  \end{figure} 
}

\def\FigGNLC{
  \begin{figure}
    \centering
    \includegraphics[width=1.0\columnwidth]{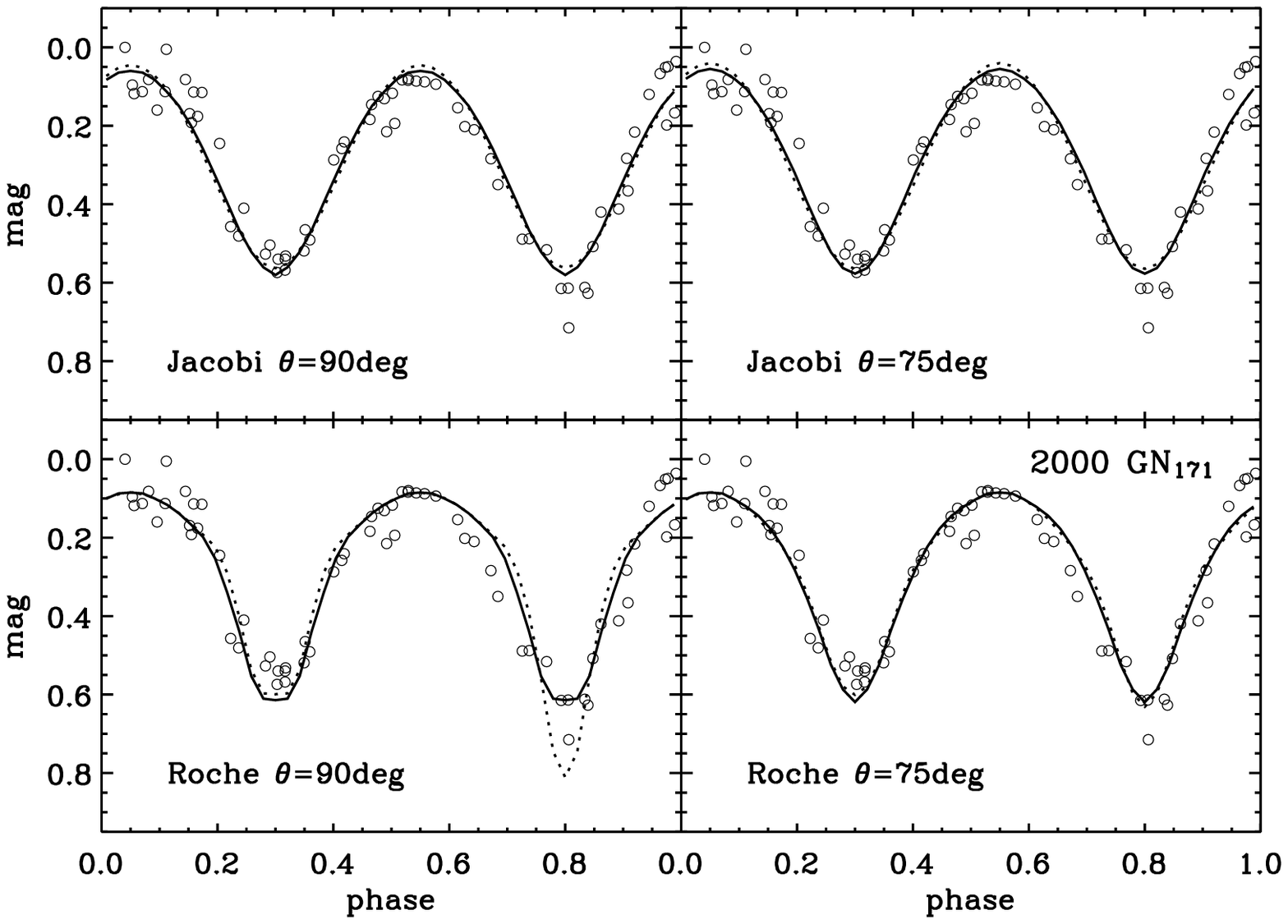}
    \caption[f12.eps] {Same as Fig.~\ref{Fig.QGLC} but for 2000$\,$GN$_{171}$.} 
	\label{Fig.GNLC}
  \end{figure}
}

\def\FigGNSh{
  \begin{figure}
    \centering
    \includegraphics[width=1.0\columnwidth]{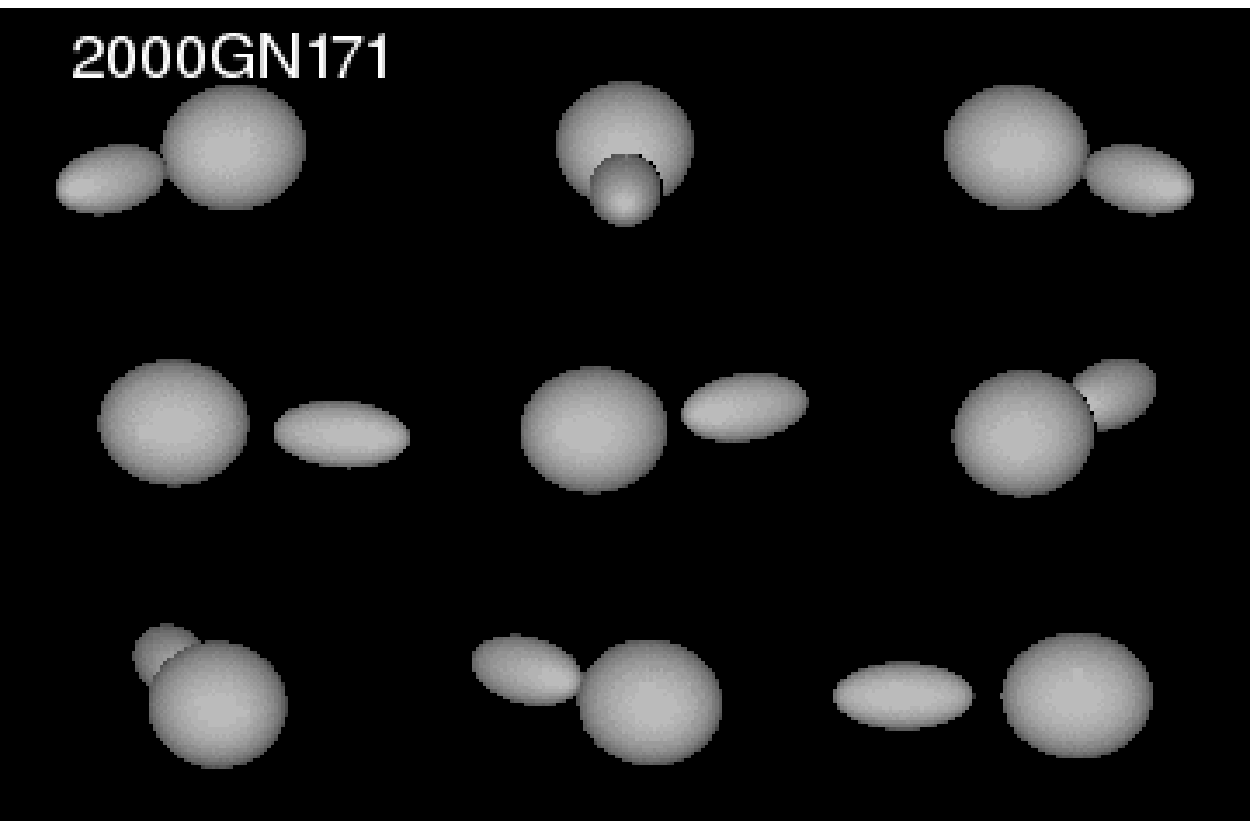}
    \caption[f13.eps] {Same as Fig.~\ref{Fig.QGSh} but for 2000$\,$GN$_{171}$.} 
	\label{Fig.GNSh}
  \end{figure}
}

\def\FigELPm{
  \begin{figure}
    \centering
    \includegraphics[width=1.0\columnwidth]{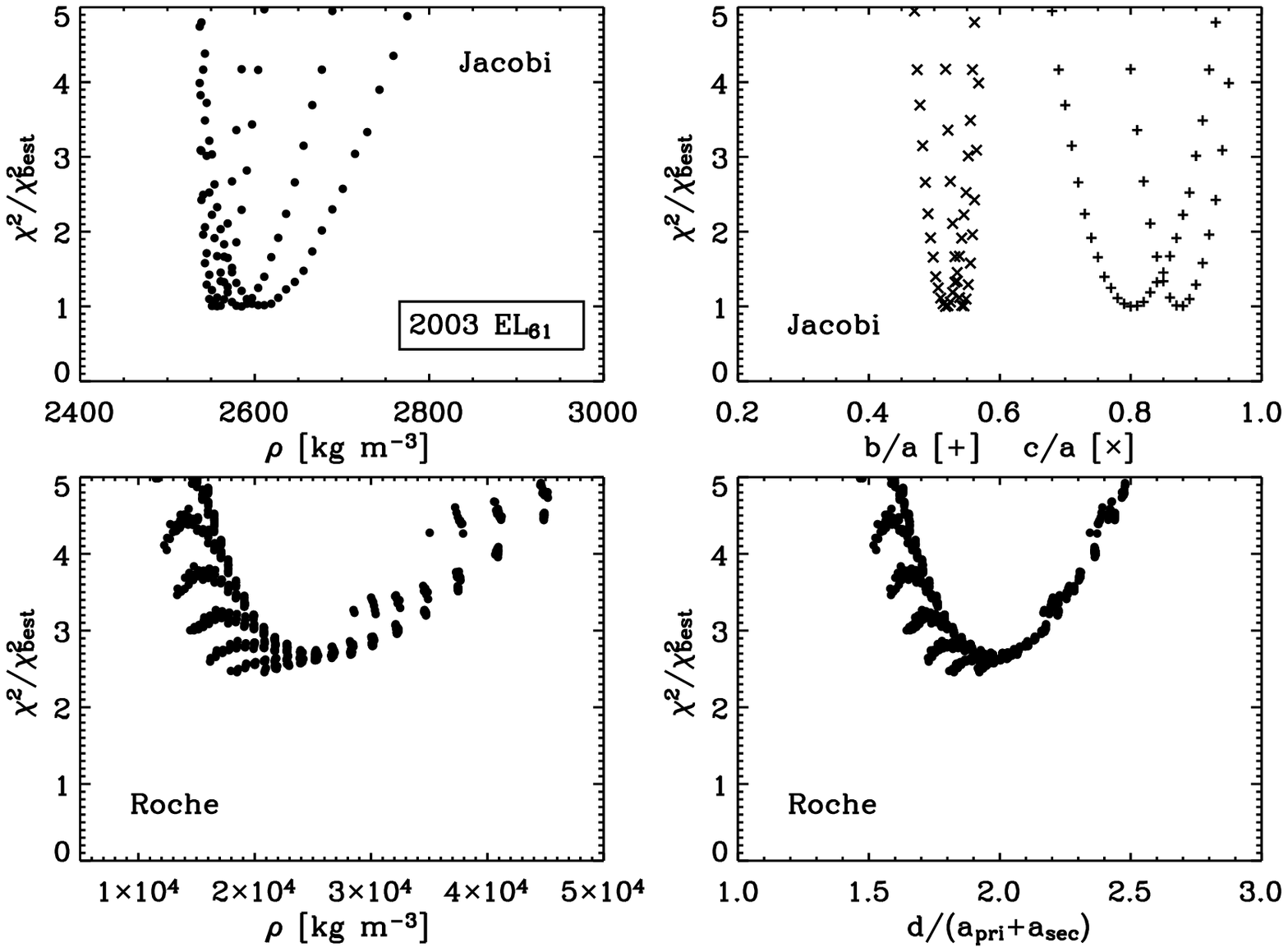}
    \caption[f14.eps] {Same as Fig.~\ref{Fig.QGPm} but for 2003$\,$EL$_{61}$ lightcurve data. Jacobi ellipsoid models are plotted for all four combinations of surface properties and observational geometry listed in Table~\ref{Table.EL}, while Roche binary models are plotted for icy-type surface at an aspect angle $\theta=75^\circ$.} 
	\label{Fig.ELPm}
  \end{figure}
}

\def\FigELLC{
  \begin{figure}
    \centering
    \includegraphics[width=1.0\columnwidth]{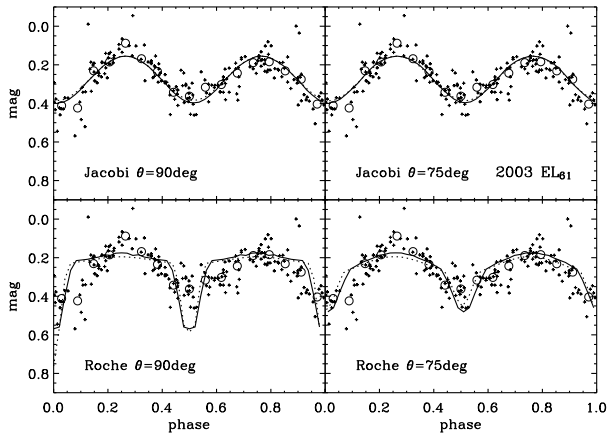}
    \caption[f15.eps] {Same as Fig.~\ref{Fig.QGLC} but for 2003$\,$EL$_{61}$.} 
	\label{Fig.ELLC}
  \end{figure}
}

\def\FigELSh{
  \begin{figure}
    \centering
    \includegraphics[width=1.0\columnwidth]{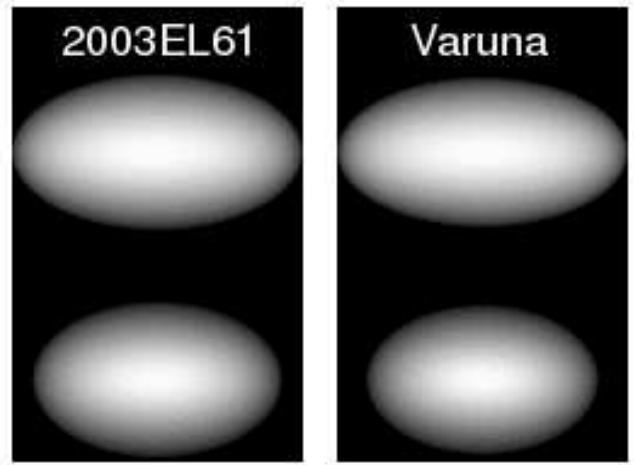}
    \caption[f16.eps] {Side (top) and tip (bottom) views of the Jacobi ellipsoid models of 2003$\,$EL$_{61}$ (left) and (20000) Varuna (right).} 
	\label{Fig.ELSh}
  \end{figure}
}

\def\FigVaPm{
  \begin{figure}
    \centering
    \includegraphics[width=1.0\columnwidth]{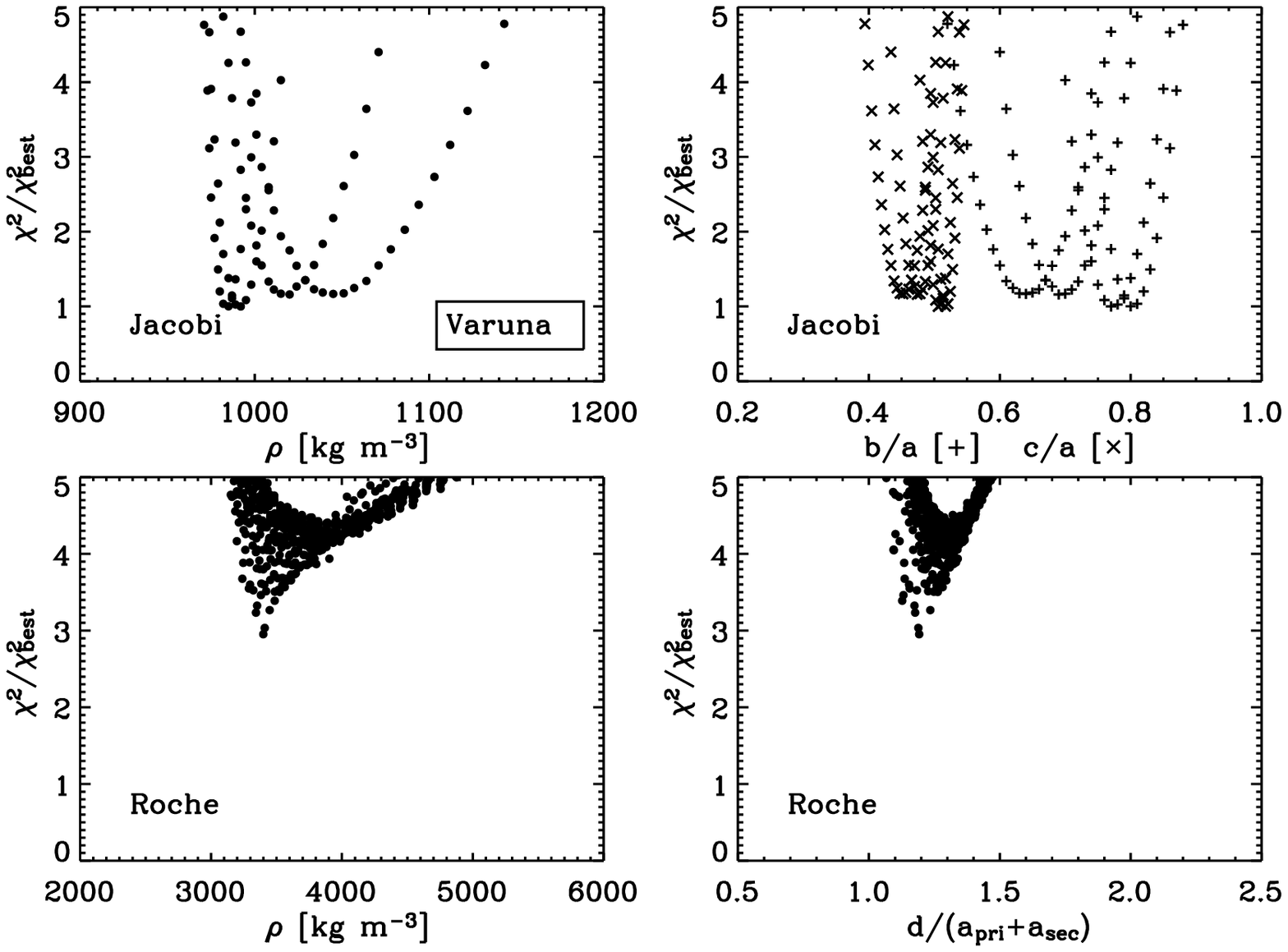}
    \caption[f17.eps] {Same as Fig.~\ref{Fig.QGPm} but for (20000) Varuna lightcurve data. Jacobi ellipsoid models are plotted for all four combinations of surface properties and observational geometry listed in Table~\ref{Table.Va}, while Roche binary models are plotted for lunar-type surface and aspect angle $\theta=75^\circ$.} 
	\label{Fig.VaPm}
  \end{figure}
}

\def\FigVaLC{
  \begin{figure}
    \centering
    \includegraphics[width=1.0\columnwidth]{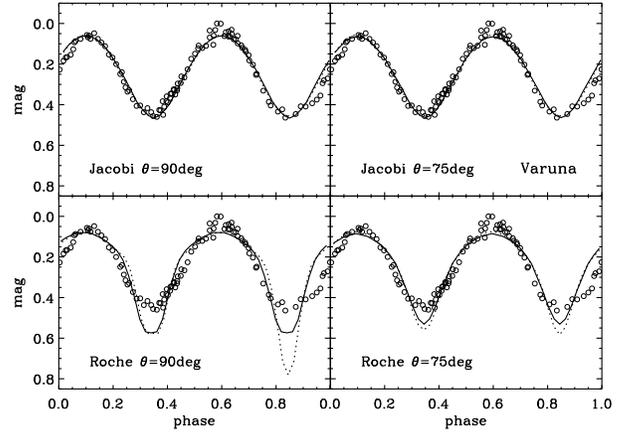}
    \caption[f18.eps] {Same as Fig.~\ref{Fig.QGLC} but for (20000) Varuna.} 
	\label{Fig.VaLC}
  \end{figure}
}

\def\FigRvsD{
  \begin{figure}
    \centering
    \includegraphics[width=1.0\columnwidth]{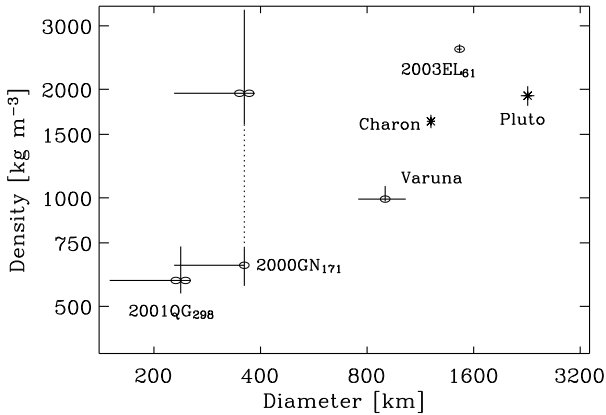}
    \caption[f19.eps] {Log density versus log equivalent circular diameter for the four KBOs modeled. Jacobi ellipsoid (Roche binary) fits are indicated by single (double) ellipsoid symbols. Pluto and Charon are plotted for comparison. KBO 2000$\,$GN$_{171}$ is plotted twice (connected by a dotted line). References in the text.} 
	\label{Fig.RvsD}
  \end{figure}
}

\def\FigGeoCorr{
  \begin{figure}
    \centering
    \includegraphics[width=1.0\columnwidth]{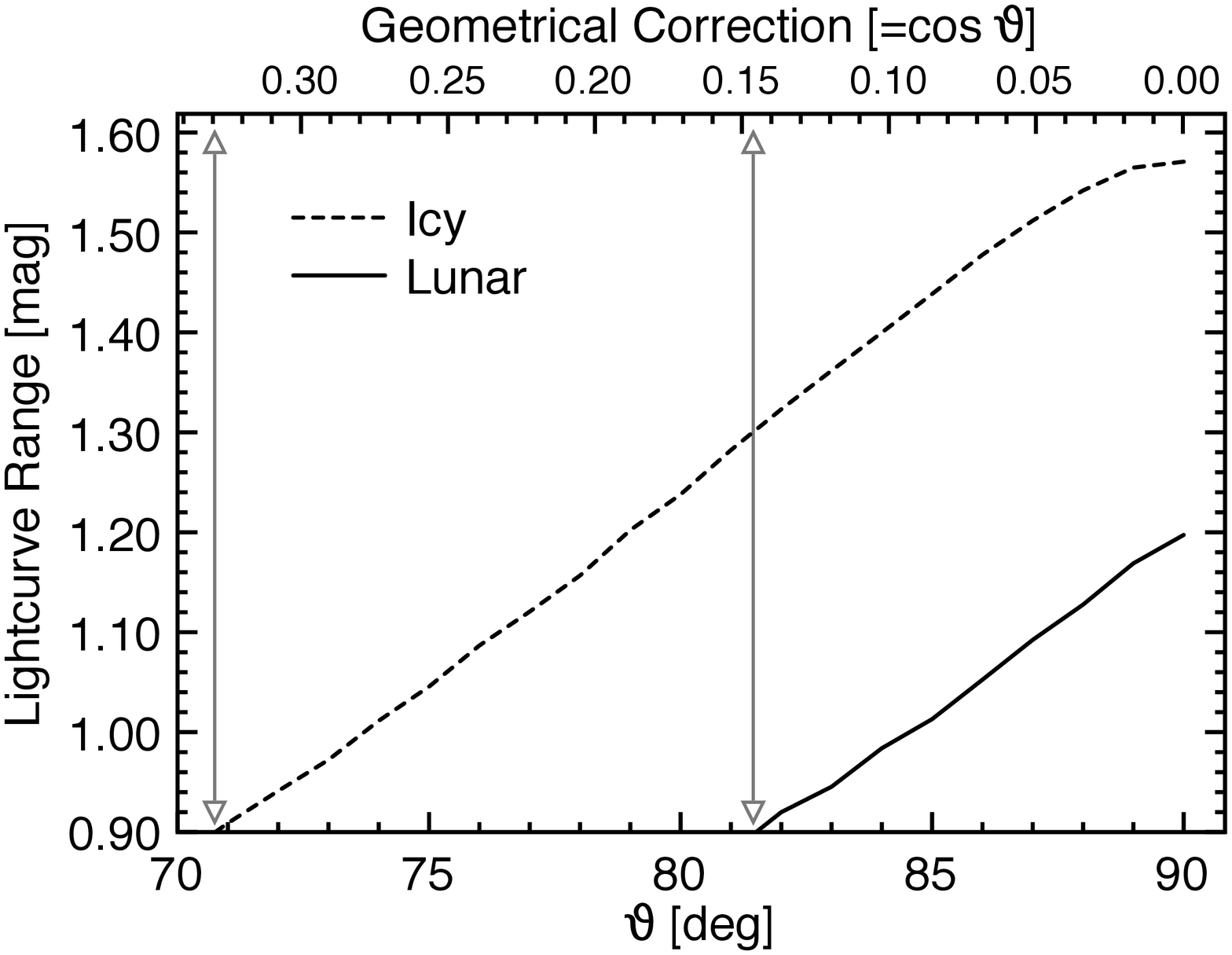}
    \caption[f20.eps] {Lightcurve range as function of aspect angle $\theta$ for maximal $\Delta m$ Roche binaries with both lunar- and icy-type surfaces. Top $x$-axis shows probability that the binary is observed at equal or larger $\theta$. See text for details.} 
	\label{Fig.GeoCorr}
  \end{figure}
}

\title{Densities of Solar System Objects from their Rotational Lightcurves}

\slugcomment{AJ, accepted 2006/12/1}
\journalinfo{AJ, accepted 2006/12/1}

\author{Pedro Lacerda\altaffilmark{1}}
\author{David C. Jewitt}
\affil{Institute for Astronomy, University of Hawaii, 2680 Woodlawn
Drive, Honolulu, HI 96822}

\email{\myemail,jewitt@ifa.hawaii.edu}

\altaffiltext{1}{GAUC, Departamento de Matem\'atica, Lg. D. Dinis, 3000 Coimbra,
Portugal}

\begin{abstract}

We present models of the shapes of four Kuiper belt objects (KBOs) and Jovian
Trojan (624) Hektor as ellipsoidal figures of equilibrium and Roche binaries.
Our simulations select those figures of equilibrium whose lightcurves best
match the measured rotational data. The best fit shapes, combined with the
knowledge of the spin period of the objects provide estimates of the bulk
densities of these objects. We find that the lightcurves of KBOs (20000) Varuna
and 2003$\,$EL$_{61}$ are well matched by Jacobi triaxial ellipsoid models with
bulk densities 992$_{-15}^{+86}\,$kg$\,$m$^{-3}$ and
2551$_{-10}^{+115}\,$kg$\,$m$^{-3}$, respectively.  The lightcurves of (624)
Hektor and KBO 2001 QG$_{298}$ are well-described by Roche contact binary
models with densities 2480$_{-80}^{+292}\,$kg$\,$m$^{-3}$ and
590$_{-47}^{+143}\,$kg$\,$m$^{-3}$, respectively. The nature of
2000$\,$GN$_{171}$ remains unclear: Roche binary and Jacobi ellipsoid fits to
this KBO are equivalent, but predict different densities, $\sim$2000\kgm and
$\sim$650\kgm, respectively.  Our density estimates suggest a trend of
increasing density with size.  

\end{abstract}

\keywords{Kuiper Belt objects --- minor planets, asteroids --- solar
  system: general}

\section{Introduction}

Most small bodies of the solar system appear unresolved at the $\sim$0.05
arcsecond peak angular resolution offered by current technology.  As a
consequence, information about the shapes and rotations of the small bodies
must be inferred, principally from measurements of the time dependence of the
scattered radiation.   So-called ``lightcurve inversion'' techniques have been
used for decades to study the rotational properties of main-belt asteroids
\citep{1989Icar...78..298C,2002aste.conf..139K}.  At their simplest, these
involve using the peak-to-peak interval of the lightcurve to estimate the
rotation period and the peak-to-peak brightness variation to estimate the
``axis ratio'' of bodies that are assumed to be triaxial in shape and in
principal axis rotation about the minor axis.  At their most complex, the
inversion techniques can be used to solve for the full 3-dimensional shapes and
rotation vectors, utilizing observations from a range of aspect angles
\citep{2001Icar..153...24K,2001Icar..153...37K}.

All lightcurve interpretations are subject to an ambiguity between variations
caused by shape and variations caused by non-uniform surface albedo, as clearly
expressed a century ago by \citet{1906ApJ....24....1R}.  This ambiguity can be
broken when simultaneous optical \textit{and} thermal observations are
available, as is the case for some of the larger asteroids in the main-belt.
Numerous observations of this type have shown that, with rare exceptions, the
albedos of the asteroids do not vary over their surfaces by a large amount
\citep{1979Icar...40..364D}.  This spatial uniformity could simply mean that
the compositions are intrinsically uniform.  Alternatively, real surface
compositional variations could exist but be smoothed-out by efficient lateral
transport of dust over the surfaces of small bodies.   The most famous
exception to this rule is provided by Saturn's satellite Iapetus, which has a
lightcurve range of nearly two magnitudes caused by surface albedo markings
\citep{1977Icar...31...81M}.  This case is pathological, however, in the sense
that it appears to be a result of Iapetus' synchronous rotation about Saturn,
which leads to unequal radiation and micrometeorite bombardment fluxes on the
leading and trailing hemispheres of the satellite.  This special geometric
circumstance is presumably not relevant to the case of small bodies in
heliocentric orbit.  

In the outer solar system, lower temperatures and greater distances make the
detection of thermal radiation increasingly challenging, even with the most
sensitive infrared satellites in space \citep[e.g.][]{2005AdSpR..36.1070C}.
Consequently, only the reflected lightcurve is available, and the
interpretation must be based on the \textit{assumption} that the surface albedo
variation is minimal.  As we will see, support for this assumption comes not
only from the analogy with the (generally uniform) main-belt asteroids, but
from the remarkably symmetric lightcurves displayed by most outer solar system
objects.  Rotational symmetry is expected for figures of equilibrium having
uniform surface albedos but is not a natural consequence of surface albedo
markings. 

The lightcurves of several large Kuiper Belt Objects (KBOs), notably (20000)
Varuna \citep{2002AJ....123.2110J} and 2003 EL61 \citep{2006ApJ...639.1238R},
suggest that these are high-angular momentum bodies in which the shape has been
deformed by rapid rotation.  Other objects may be contact binary systems, as
has long been suggested for Jovian Trojan (624) Hektor
\citep{1978Icar...36..353H,1980Sci...207..976H,1980Icar...44..807W} and,
recently, for KBO 2001$\,$QG$_{298}$
\citep{2004AJ....127.3023S,2004PASJ...56.1099T}.  These systems are interesting
since, under conditions of rotational equilibrium, the period and the shape
(both of which can be inferred from lightcurve data) are uniquely related to
the bulk density.  Lightcurves of these objects may thus be interpreted in
terms of a fundamental geophysical property that is otherwise difficult to
measure.  

\input{tab1.tex}

\input{tab2.tex}

In this paper, we discuss the lightcurves of specific solar system bodies in
terms of rotational equilibrium models, paying particular attention to high
angular momentum systems and contact binaries.  Our models address the effects
of the surface scattering on the derived system parameters.  Prototype contact
binary (624) Hektor is examined in detail, taking advantage of voluminous high
quality data published for this object over a range of aspect angles (see
Table~\ref{Table.HektorObs}). The models are then applied to four well-observed
KBOs (Table~\ref{Table.BodyList}) and used to place quantitative contraints on
their properties in a consistent formalism. Indeed, the uniformity of approach
is one of the strengths of our simulations.

\section{Lightcurve Simulations}
\label{Sec.LS}

\subsection{Jacobi Ellipsoids}
\label{SubSec.TFoE}

The formalism associated with the ellipsoidal figures of equilibrium is
described in great detail in \citet{1969efe..book.....C}. A homogeneous, fluid
body spinning in free space will assume a shape that balances self-gravity and
the inertial acceleration due to rotation. This means that the triaxial shape
of such a body is a function of its spin frequency and density.  The
equilibrium figures of isolated, rotating bodies are the Maclaurin spheroids
and the Jacobi ellipsoids. The former are oblate spheroids and the latter are
triaxial ellipsoids, and in both cases the rotation is about the shortest
physical axis.  We are interested only in the Jacobi ellipsoids because oblate
spheroids have rotational symmetry and thus produce flat lightcurves. 

The shapes of Jacobi ellipsoids in terms of the semi-axes $(a,b,c)$ can be
obtained by solving \citep{1969efe..book.....C}
\begin{eqnarray} 
a^2 b^2 \int_{0}^{\infty}\frac{1}{\left(a^2+u\right)\left(b^2+u\right)\,\Delta(a,b,c)}\,du = \\ \nonumber
c^2\int_{0}^{\infty}\frac{1}{\left(c^2+u\right)\,\Delta(a,b,c)}\,du
\label{EqTriaxialShape}
\end{eqnarray}
\noindent where
$\Delta(a,b,c)=\sqrt{\left(a^2+u\right)\left(b^2+u\right)\left(c^2+u\right)}$.
The spin frequency $\omega$ and density $\rho$ are related to the shape by
\begin{equation}
\frac{\omega^2}{\pi\,G\,\rho}=2\,a\,b\,c\int_{0}^{\infty}\frac{u}{\left(a^2+u\right)\left(b^2+u\right)}\,du
\label{EqTriaxialOmegaRho}
\end{equation} \noindent where $G$ is the gravitational constant. We solved
Eq.~(\ref{EqTriaxialShape}) for values of $b/a$ between 0.43 and 1.00 in steps
of 0.01, and used the solutions, together with Eq.~(\ref{EqTriaxialOmegaRho}),
to calculate $\omega^2/(\pi\,G\,\rho)$ which relates the spin period to the
body density for each of the equilibrium triaxial ellipsoids. Figures with
$b/a<0.43$ are unstable to rotational fission \citep{1919QB981.J4.......}.

The derived shapes are then raytraced at regular intervals spanning a full
rotation period.  This produces a set of frames from which the lightcurve is
extracted by integrating the total light in each one of them. In this way we
generate a database of lightcurves of figures of equilibrium which can be used
to compare to the lightcurve data. As described below we run our simulations
for two surface scattering laws.

\FigAvsN

The raytracing is done using the main engine of the open source software
POV-Ray (http://www.povray.org). The surface scattering routines were rewritten
to permit accurate control of the scattering function.  To test the accuracy of
our raytracing method we simulated the lightcurve of a triaxial ellipsoid with
an axis ratio $b/a$, observed equator-on ($\theta = 0^\circ$) at zero phase
angle ($\alpha=0^\circ$), using a ``lunar'' surface scattering function (see
Section~\ref{SubSec.SS}), and compared the result with the analytical solution
for the same configuration, given by
\begin{equation}
m(b/a,\phi)=2.5\,\log_{10} \sqrt{1 + \left[ (b/a)^2 - 1\right] \,{\cos^2 (2\,\pi \,\phi )}}.
\end{equation}
\noindent where $\phi\in[0,1]$ is the rotational phase of the ellipsoid. The
result is plotted in Fig.~\ref{Fig.AvsN}. The raytraced lightcurve deviates
$\sim0.1\%$ from the analytical solution which is negligible when compared to
the uncertainties in the photometric data, typically $\sim2$ to $3\%$ on real
astronomical objects.

\subsection{Roche Binaries}
\label{SubSec.Binaries}

Lightcurves from an eclipsing binary asteroid consisting of two spheres in
circular orbit have been presented by \citet{1979Icar...40..383W}.
As the binary separation becomes comparable to the scale of either component,
mutual gravitational forces will deform the bodies, increasing the lightcurve
range over the maximum (factor of two) possible for equal-sized spheres
\citep{1984A&A...140..265L}.
To model this mutual deformation in close and contact binary systems we use the
Roche binary approximation \citep{1963ApJ...138.1182C,1984A&A...140..265L}. In
this approximation each component is considered to be a Roche ellipsoid, which
is the equillibrium shape of a satellite orbiting a spherical, more massive
primary. The tidally deformed shape of the secondary is assumed to be solely
caused by the spherically symmetric gravitational gradient due to the primary.
Each component's shape is calculated separately using reciprocal values, $q$
and $1/q$, for the mass ratio. Clearly, such an approximation introduces the
most error when calculating shapes of close binaries with mass ratios near
$q=1$. In these situations the elongation of the binary components is
underestimated which leads to smaller lightcurve ranges. With the further
assumptions that the binary is tidally locked, and that the components have
equal density $\rho$, and orbit the center mass in circular paths, the mass
ratio $q$, the shape of one of the components $(b/a,c/a)$, and the orbital
frequency $\omega$ can be calculated by solving \citep{1963ApJ...138.1182C}
\begin{subequations}
\begin{gather} \label{EqRocheEllA}
\frac{(3+1/q)\,a^2+c^2}{(1/q)\,b^2+c^2} = \frac{a^2 A_1-c^2 A_3}{b^2 A_2-c^2 A_3} \\
\label{EqRocheEllB} \frac{q}{1+q}\,\frac{\omega^2}{\pi\,G\,\rho} = 2\,a\,b\,c\,\frac{a^2 A_1-c^2 A_3}{(3+1/q)\,a^2+c^2},
\end{gather}
\end{subequations}
\noindent with $A_1$, $A_2$, and $A_3$ given by \citep{1962ApJ...136.1037C}
\begin{subequations}
\begin{align}
A_1 &= \frac{2}{a^3 \sin^3\phi}\,\frac{1}{\sin^2\theta}\left[F(\theta,\phi)-E(\theta,\phi)\right] \\
A_2 &= \frac{2}{a^3 \sin^3\phi}\,\frac{1}{\sin^2\theta \cos^2\theta} \times \\ \nonumber
 & \left[E(\theta,\phi)-F(\theta,\phi)\cos^2\theta-\left(\frac{c}{b}\right)\sin^2\theta\sin\phi\right] \\
A_3 &= \frac{2}{a^3 \sin^3\phi}\,\frac{1}{\cos^2\theta}\left[\left(\frac{b}{c}\right)\sin\phi-E(\theta,\phi)\right],
\end{align}
\end{subequations}
\noindent where $E(\theta,\phi)$ and $F(\theta,\phi)$ are the standard elliptic
integrals of the two kinds with arguments
\begin{equation}
\theta = \arcsin\sqrt\frac{a^2-b^2}{a^2-c^2}\quad \mathrm{and}\quad \phi = \arccos \left(\frac{c}{a}\right).
\end{equation}

Each root of Eq.(\ref{EqRocheEllA}), corresponding to a Roche ellipsoid
solution, can be calculated by setting 3 of the 4 parameters $q$ and $(a,b,c)$,
and solving for the 4th by interpolation. For the primary, we set $a=1$ and
calculate $b$ for each combination of $q=q_\textrm{min},\dots,1.00$ and
$c=c_\textrm{min},\dots,0.99$, both in steps of 0.01.  The procedure is
repeated using $1/q$ instead of $q$ to calculate the shape of the secondary,
$(a',b',c')$. A valid Roche binary solution is obtained if two sets,
$(q,a,b,c)$ and $(1/q,a',b',c')$, yield the same value
$\omega^2(\pi\,G\,\rho)^{-1}$ when replaced into the right-hand side of
Eq.(\ref{EqRocheEllB}). Table~\ref{Table.RocheEll} shows the solutions for
$q=0.25$.

\input{tab3.tex}

\subsection{Surface Scattering}
\label{SubSec.SS}

The surface scattering properties of KBOs are unknown. For this reason the
amount of sunlight reflected from a KBO is usually taken to be proportional to
its geometrical cross-section. However, the total range of the lightcurve of a
convex object increases significantly if there is limb darkening. The two
simplest scattering laws generally used to model planetary surfaces are the
Lommel-Seeliger and Lambert laws. The Lommel-Seeliger law is a single
scattering model, suitable for low albedo surfaces, and can be considered a
simplification of the well known Hapke model \citep{2005PhDT.........7L}. The
Hapke model is inappropriate for this work because it has many parameters which
cannot realistically be constrained using lightcurve data. To model relative
brightness variations, which is what is needed to generate lightcurves, the
Lommel-Seeliger law requires no parameters: it depends solely on the cosines of
the incidence and emission angles ($i$ and $e$, the angles between the surface
normal and the directions to the light source and to the observer). The
Lommel-Seeliger reflectance function is thus 
\begin{equation}
r_\mathrm{LS} \propto\frac{\mu_0}{\mu+\mu_0} 
\end{equation}
where $\mu_0=\cos i$ and $\mu=\cos e$.  

The Lambert scattering law is a simple description of a perfectly diffuse
surface. It assumes that a light ray that enters the material is multiply
scattered and thus leaves the surface in a random direction. As such, it is a
multiple scattering law which adequately describes high albedo surfaces.  The
Lambert reflectance function is 
\begin{equation} 
r_\mathrm{L} \propto\mu_0.  
\end{equation}
The Lommel-Seeliger and Lambert scattering functions are taken
here as representative of low albedo ``lunar''-type surfaces and high albedo
``icy''-type surfaces, respectively.

\FigLCurSLawPAngEq
\FigLCurSLawPAngUn
\FigLCurSLawPAngAr

Figures \ref{Fig.LCurSLawPAngEq} and \ref{Fig.LCurSLawPAngUn} compare the
lightcurves of Roche contact binaries using both lunar and icy scattering
models, at four different phase angles. A ``uniform'' scattering law is also
plotted for comparison. The uniform model assigns equal brightness to any
illuminated point on the surface, and thus represents the illuminated
cross-section.  Figure \ref{Fig.LCurSLawPAngEq} is for a contact binary with
equal size components, while Fig. \ref{Fig.LCurSLawPAngUn} represents the case
of different sizes for primary and secondary (see Fig.
\ref{Fig.LCurSLawPAngUn}).  A few conclusions can immediately be drawn from
inspection of Figs.  \ref{Fig.LCurSLawPAngEq} and \ref{Fig.LCurSLawPAngUn}:
\begin{itemize} 

\item At low phase angle the lunar model produces negligible limb darkening and
is thus equivalent to the simpler uniform model. Only at large phase angles
($\gtrsim 30\,$deg) does the uniform approximation fail to follow the lunar
scattering law.  Icy surfaces, however, always produce larger lightcurve ranges
for the same shape, which implies that assuming uniform scattering when
interpreting lightcurves of icy objects will tend to exaggerate the inferred
shape elongation.

\item The lightcurve minima become broader with increasing phase angle. This
fact implies that V-shaped minima are strictly diagnostic of a close binary
configuration only when viewed near zero phase angle. Fortunately, this is
necessarily the case for all observations of Kuiper belt objects. 

\item Observations at different phase angles shift the minima and maxima of the
lightcurve in rotational phase. This effect should be taken into account when
fitting a single spin period to observations taken at different phase angles.
For instance, failure to fit all the data with a single period does not
necessarily imply complex (non-principal axis) rotation when observations over
a wide range of phase angles are compared.  

\end{itemize}

In our simulations we use the lunar- and icy-type surface laws separately to
assess how different surface properties affect our results. As shown below, we
find that different surface properties do not significantly change our density
estimates.  It is also shown that while in some cases the choice of a
particular scattering law clearly improves the lightcurve fit, in other cases
the result is degenerate as far as surface properties are concerned. The
simulations further assume that the surface albedo is uniform and the same for
both components of the binary.

\subsection{Observational Geometry}
\label{SubSec.Geometry}

Owing to their large distances from the Sun and the Earth, KBOs can only be
observed at small phase angles ($\alpha < 2\,$deg). (20000) Varuna is known to
exhibit a pronounced opposition effect at phase angle $\alpha < 0.1\,$deg
\citep{2005Icar..176..492H} which seems to affect the extent of its brightness
variation \citep{2006Icar..184..277B}. In the range
$0.1\,\mathrm{deg}<\alpha<2.0\,$deg the phase curve of (20000) Varuna is linear
\citep{2002AJ....124.1757S} and the lightcurve unnaffected by phase effects.
We place our simulations well within this linear regime by using a phase angle
$\alpha=1\,$deg. In addition, since most of the data being modeled have
$\alpha>0.1\,$deg the opposition effect is likely unimportant to the
conclusions presented here.

The aspect angle is the angle between the line-of-sight and the spin axis of
the system. We simulated lightcurves for two values of the aspect angle: 90
degrees (equator-on) and 75 degrees. Given that the spin axis orientations are
unknown, the models with a tilted orientation allow us to investigate how the
aspect angle affects our conclusions. In the case of (624) Hektor the spin
(orbital) axis orientation is known \citep{1969AJ.....74..796D} and we were
able to perform simulations using the actual aspect and phase angles at the
moment the data were taken (see Table~\ref{Table.HektorObs}).

\section{Best Fit Solutions}

\subsection{(624) Hektor}
\label{SubSec.Hektor}

\input{tab4.tex}
\FigHePm
\FigHeLC
\FigHeSh

Jovian Trojan (624) Hektor has long been recognized as a likely contact binary
\citep{1978Icar...36..353H,1980Sci...207..976H,1980Icar...44..807W}. Numerous
lightcurve data spanning a long time base (1957--1968) have been collected for
this object, which allowed the determination of the pole orientation
\citep{1969AJ.....74..796D,1995P&SS...43..649D}.  Depending on the orbital
configuration, the lightcurve range of (624) Hektor varies between 0.1$\,$mag
and 1.2$\,$mag (see Table~\ref{Table.HektorObs}). As noted by \citet[see also
\citealt{1984A&A...140..265L}]{1980Icar...44..807W}, lightcurve ranges above
0.9$\,$mag cannot be produced by rotation of a single equilibrium figure and,
instead, are most likely produced by a tidally distorted contact binary.
Besides the large range of variability, the lightcurve morphology of (624)
Hektor exhibits V-shaped minima and round maxima, also characteristic of
tidally deformed, contact binary systems.

Making use of the extensive dataset presented in \citet{1969AJ.....74..796D},
we have selected the Roche binary model (see Section~\ref{SubSec.Binaries})
that simultaneously best fits the observations at four observing campaigns (see
Table~\ref{Table.HektorObs}).  The quality of fit is measured by
$\chi^2/\chi^2_\mathrm{best}$, i.e., the ratio of the chi-squared value of each
model to the chi-squared value of the best fit model. This corresponds to a
reduced chi-squared, $\chi^2_\mathrm{red}$, if one assumes that the best fit
model has $\chi^2_\mathrm{red}=1$. We do this because the errors associated
with the data for (624) Hektor are not known with certainty, which does not
allow us to reliably calculate the reduced chi-squared for each model.   In
Table~\ref{Table.HektorFit} we show the three Roche models that best
approximate (624) Hektor's lightcurve. Figure~\ref{Fig.HePm} shows how well our
simulations are able to determine the density, orbital distance, mass ratio,
and surface scattering properties of the (624) Hektor system.  In the upper
left panel we plot histograms of quality of fit for the two scattering laws
considered. A lunar scattering law clearly produces better fits, which is no
surprise as (624) Hektor is known to have a low albedo
\citep[$\sim$0.06,][]{1977Icar...30..224C,2003AJ....126.1563F}. In the lower
left panel the orbital distance of the Roche binary models is plotted versus
quality of fit.  The orbital distance is given in units of the sum of the
semi-major axes of the primary, $A$, and of the secondary, $a$.  The minimum is
centered around $d/(A+a)=1$, which corresponds to the binary components being
in contact.  Values $d/(A+a)<1$ are unphysical but we have decided to keep them
for two reasons. Firstly, because they may result from the approximations of
the Roche model \citep{1984A&A...140..265L}.  The orbital distance is
calculated using Kepler's 3rd law assuming point masses for the binary
components. Since these have elongated shapes, gravity will be enhanced meaning
the Roche model may underestimate the orbital distance.  Secondly, because the
mechanism that brought the two components together and formed the binary may
have produced some deformation (a ``crush'' zone) around the point of contact,
bringing the objects closer together than the situation of contact between two
perfectly hard ellipsoids. Considering models with
$\chi^2/\chi^2_\mathrm{best}<2$, which roughly corresponds to a 1-$\sigma$
criterion, we find that the two components of (624) Hektor are separated by
$d/(A+a)=1.00^{+0.11}_{-0.09}$, which is consistent with contact. The top right
panel of Fig.~\ref{Fig.HePm} shows that the model poorly constrains the mass
ratio of the binary. The mass ratio which corresponds to the best fit Roche
binary is $q=0.62$, but the range of $q$ values that fall within 1-$\sigma$ of
the best fit is broad.  As for the bulk density of (624) Hektor (bottom right
panel of Fig.~\ref{Fig.HePm}), perhaps the most interesting quantity to come
out of our simulations, we find $\rho=2480^{+80}_{-292}\,$kg$\,$m$^{-3}$. This
value closely confirms an earlier estimate: $\rho\sim2500\,$kg$\,$m$^{-3}$
\citep{1980Icar...44..807W}.

Figure~\ref{Fig.HeLC} shows how the best Roche binary model compares to the
\citet{1969AJ.....74..796D} lightcurve data. The differences in lightcurve
range from one campaign to the next reflect the effect of the observational
geometry. Given the simplicity of the model, the agreement is remarkable and
lends strong support to the idea that (624) Hektor is a contact binary. The
model for Julian day 2438795 shows the largest departure from the data. This is
to be expected given the small aspect angle, $\theta=24.8\,$deg.  The
cross-section of the binary varies little at such unfavorable geometry and
brightness variations must be attributed to irregularities on the surface of
the Trojan, which are not accounted for in the model. In Fig.~\ref{Fig.HeSh} we
show the best fit model rendered at the four aspect angles and for eight values
of rotational phase.  Recent observations using the LGS AO system at the
Keck-II telescope suggest that (624) Hektor may have a bilobated shape
\citep{2006IAUC.8732....1M} and lend further support to the results presented
here.

\subsection{2001$\,$QG$_{298}$}
\label{SubSec.QG298}

\input{tab5.tex}
\FigQGPm
\FigQGLC
\FigQGSh

Kuiper belt object 2001$\,$QG$_{298}$ completes a full rotation every
$P=13.77\,$hr and its brightness varies by $\Delta m=1.14\pm 0.04\,$mag
\citep{2004AJ....127.3023S}. The large range of brightness variation and
relatively slow rotation provide compelling evidence that 2001$\,$QG$_{298}$ is
a contact or very close binary \citep{2004AJ....127.3023S}.  We attempted to
fit both Jacobi ellipsoid and Roche binary models to the lightcurve data on
2001$\,$QG$_{298}$, using the two surface scattering laws and the two different
observational geometries described in Sections~\ref{SubSec.SS} and
\ref{SubSec.Geometry}.  Table~\ref{Table.QG} and Figures~\ref{Fig.QGPm} and
\ref{Fig.QGLC} present a summary of the best fit models for different
combinations of scattering law and observational geometry. 

Clearly, the Roche binary simulations fit the data better. Furthermore, the
binary model favored by the data has a lunar-type surface and an equator-on
geometry (see Fig.~\ref{Fig.QGLC}). However, choosing an icy-type surface does
not result in a significantly poorer fit. Indeed, the lower left panel of
Fig.~\ref{Fig.QGLC} suggests that an intermediate scattering law is needed to
fit the shallower minimum in the lightcurve data.  Models tilted 15 degrees
toward the line of sight are unable to fit the deeper V-shaped minimum in the
data.  Taking all the Roche simulations into account we find that
2001$\,$QG$_{298}$ should have an orbital separation
$d/(A+a)=0.90_{-0.14}^{+0.31}$ (contact binary) and a bulk density
$\rho=590_{-47}^{+143}\,$kg$\,$m$^{-3}$. The uncertainty intervals are
established in the same way as was done for (624) Hektor (see
Section~\ref{SubSec.Hektor}).  Inspection of Table~\ref{Table.QG} and
Fig.~\ref{Fig.QGPm} shows that the best icy-type surface models have
$d\sim1.09$ and $\rho\sim660\,$kg$\,$m$^{-3}$. Given that the chosen surface
scattering laws represent extreme cases (the surface of 2001$\,$QG$_{298}$
probably combines single and multiple scattering behaviour), we must conclude
that the density we find does not depend strongly on the specific surface
scattering properties of the KBO.  The same applies to the binary components
being in (or very close to) contact. The best fit Roche binary is rendered in
Fig.~\ref{Fig.QGSh}.  Our results are in good agreement with an independent but
similar attempt to fit this object's lightcurve with a Roche binary model
\citep{2004PASJ...56.1099T}.

\subsection{2000$\,$GN$_{171}$}
\label{SubSec.GN171}

\input{tab6.tex}
\FigGNPm
\FigGNLC
\FigGNSh

The rotational properties of KBO 2000$\,$GN$_{171}$ also make it a good
candidate contact binary \citep{2004AJ....127.3023S}.  Its spin period and
lightcurve range are $P=8.329\,$hr and $\Delta m=0.61\pm 0.03\,$mag
\citep{2002AJ....124.1757S}. However, as can be seen in Table~\ref{Table.GN}
and Figures~\ref{Fig.GNPm} and \ref{Fig.GNLC}, the lightcurve fitting results
are not definitive about the nature of this KBO. A Roche binary solution is the
one that best fits the data (see Table~\ref{Table.GN}) but it is not
significantly better than a single Jacobi ellipsoid model.  Inspection of the
lightcurve fits (Fig.~\ref{Fig.GNLC}) suggests that while the Jacobi ellipsoid
model follows better the overall shape of the lightcurve, it is not able to
reproduce the different minima present in the data. A Roche binary solution
produces different minima, but seems to require an aspect angle
$\theta<90\,$deg and a low mass ratio ($q\sim0.3$) to be able to reproduce a
lightcurve range as low as $\Delta m=0.61\,$mag. The predicted orbital
separation is $d/(A+a)=1.09_{-0.10}^{+0.55}$. Lunar-type surface models
consistently produce better fits than icy-type, irrespective of the nature
(Jacobi ellipsoid or Roche binary) and orientation of 2000$\,$GN$_{171}$.  The
inferred density is model dependent, but for each of the two models it does not
depend much on scattering properties nor on geometry.  If 2000$\,$GN$_{171}$ is
taken to be a binary, its density should be $\rho\sim2000\,$kg$\,$m$^{-3}$.
Instead, if it is an elongated ellipsoid it should have a bulk density
$\rho\sim650\,$kg$\,$m$^{-3}$. The best Roche binary model for
2000$\,$GN$_{171}$ is rendered in Fig.~\ref{Fig.GNSh}.

\subsection{2003$\,$EL$_{61}$}
\label{SubSec.EL61}

\input{tab7.tex}
\FigELPm
\FigELLC
\FigELSh

The lightcurve of 2003$\,$EL$_{61}$ indicates a rotation period of $P=3.9\,$hr
and a lightcurve total range of $\Delta m=0.28\pm0.04$
\citep{2006ApJ...639.1238R}. The extremely fast rotation of 2003$\,$EL$_{61}$
implies that it must have a high density. Using a hydrostatic equilibrium
criterion, \citet{2006ApJ...639.1238R} estimate
$\rho\sim2600-3340\,$kg$\,$m$^{-3}$. A binary solution would require a
considerably higher (and unrealistic) density than a rotationally deformed
ellipsoid. Binarity is also unlikely given the small range of brightness
variation: for a binary to produce such a shallow lightcurve, the pole axis
must nearly coincide with the line of sight. Indeed, we find that no Roche
binary model is able to satisfactorily fit this object's lightcurve data (see
Table~\ref{Table.EL} and Fig.~\ref{Fig.ELPm}).  In the case of Jacobi ellipsoid
models, all possible combinations of surface properties or orientation fit the
data equally well.  This is partly due to the large scatter present in the
lightcurve data.  However, the predicted density
($\rho=2585_{-44}^{+81}\,$kg$\,$m$^{-3}$) depends little on specific choices of
surface and geometry, and is consistent with the
$\rho=2600-3340\,$kg$\,$m$^{-3}$ estimate of \citet{2006ApJ...639.1238R}. We
find that the axis ratios of 2003$\,$EL$_{61}$ should fall in the ranges
$b/a=0.76-0.88$ and $c/a=0.50-0.55$. The icy scattering law (with
$\theta=75\,$deg; see Table~\ref{Table.EL}) is preferable, as
\citet{2006ApJ...639.1238R} find that 2001$\,$EL$_{61}$ has a high albedo
($>$0.6). The best Jacobi ellipsoid representation of 2001$\,$EL$_{61}$ is
shown in Fig.~\ref{Fig.ELSh}.

\subsection{(20000) Varuna}
\label{SubSec.Var}

\input{tab8.tex}
\FigVaPm
\FigVaLC

The rotational properties of this object ($P=6.34\,$hr and $\Delta
m=0.42\pm0.02$) were interpreted in the context of ellipsoidal figures of
equilibrium and a density of $\rho\sim1000\,$kg$\,$m$^{-3}$  was derived
\citep{2002AJ....123.2110J}. Our simulations lend support to this result by
showing that (20000) Varuna's lightcurve is well fit by a Jacobi ellipsoid
model. This is apparent from Table~\ref{Table.Va} and Figs.~\ref{Fig.VaPm} and
\ref{Fig.VaLC}. Figure~\ref{Fig.ELSh} depicts the Jacobi ellipsoid model of
Varuna. As in the case of 2003$\,$EL$_{61}$, the quality of fit is degenerate
as far as surface properties and orientation are concerned. Thus, depending on
particular choices of these properties, Varuna's axis ratios lie in the ranges
$b/a=0.63-0.80$, $c/a=0.45-0.52$. The bulk density determination is again much
more robust: we find $\rho=992_{-15}^{+86}\,$kg$\,$m$^{-3}$.

\section{Discussion}

The photometric lightcurve of 2003$\,$EL$_{61}$ exhibits asymmetries which are
not reproduced by the simple Jacobi ellipsoid model. Given the large size of
this object ($D_\mathrm{eq}\sim1450$, see Table~\ref{Table.BodyList}) which
safely puts it in the gravity regime, we do not expect such irregularities in
the lightcurve to be due to an irregular shape. Instead, if the lightcurve
features are real, they could have the same origin as Pluto's brightness
variation: albedo patches across the object's surface. Like Pluto,
2003$\,$EL$_{61}$ is large enough to hold a thin atmosphere, which might
condense on the surface and cause the patches.

The high density derived for (624) Hektor stands in contrast to the low value
($\rho$ = 800$_{-100}^{+200}$ kg m$^{-3}$) derived for resolved Trojan binary
(617) Patroclus \citep{2006Natur.439..565M}.  The sizes of these two Trojans
are similar: Hektor is 102$\pm$2 km in radius while Patroclus is 70$\pm$2 km in
radius, when measured and interpreted in the same way
\citep{2003AJ....126.1563F}.  The low density of Patroclus requires substantial
porosity and also suggests an ice-rich composition.  Hektor's density is
consistent with zero porosity and a smaller or negligible ice fraction.  This
difference is puzzling, given that the albedos (0.057$\pm$0.004 and
0.050$\pm$0.005, respectively) are very similar, as are the optical
reflectivity gradients
[11.6$\pm$1$\,$\%/1000$\,$\AA~\citep{1991PhDT.......105S} and
8.8$\pm$1$\,$\%/1000$\,$\AA~\citep{1990AJ....100..933J}, respectively].

\citet{2006Natur.439..565M} argued that the low density and inferred porous,
ice-rich composition of (617) Patroclus was an indication that it originated in
the outer part of the solar system. The high density of (624) Hektor is hard to
explain in this context; does it indicate that (624) Hektor did not form in the
outer solar system? Analogously, the similar size and surface properties of
these two Jovian Trojans could be used to infer a {\em common} origin.  A
similar density argument was used for the Saturnian irregular satellite Phoebe
($\sim220\,$km in radius), but in the opposite direction. The {\em high}
density of Phoebe ($\rho=1630\pm33\,$kg$\,$m$^{-3}$), when compared with that
of other (regular) moons of Saturn, has been intepreted as indicative of an
outer solar system origin on the basis that it matches the density of Pluto
\citep{2005Natur.435...69J}. As the examples above show, it is difficult to
establish a simple relation between formation region and bulk density -- there
may be no such relation -- and therefore the density of a body alone should not
be used to infer its origin.

\FigRvsD

Figure \ref{Fig.RvsD} shows the KBO densities from our simulations versus
equivalent circular diameter; Pluto and Charon \citep{2006AJ....132.1575P} are
plotted for comparison.  The sizes of 2001$\,$QG$_{298}$ and 2001$\,$GN$_{171}$
were calculated from their absolute magnitude assuming a 0.04 albedo, and the
error bars extend the albedo to 0.10.  The size of (20000) Varuna is from
\citet{2001Natur.411..446J} and that of 2003$\,$EL$_{61}$ is calculated using
its mass \citep{2006ApJ...639.1238R} and the density derived here (see also
Table~\ref{Table.BodyList}). A trend of increasing density with size is clear.
Such relation may be caused by (1) a difference in composition (ice/rock
ratio), with bigger objects having larger rock fractions, or (2) a trend in
porosity, with larger objects being more compacted than their smaller
counterparts, likely due to larger internal pressure. Although the latter
effect is certainly present, it is unclear if it is the dominant cause for the
trend. This size-density relation has been noted before
\citep{2002acm..conf...11J} and seems to be present in different populations,
e.g., KBOs and planetary satellites \citep{2006saas.conf.....J}.

\section{The Fraction of Contact Binaries in the Kuiper Belt}

Our simulations can be used to determine the lightcurve range of Roche binaries
at arbitrary observing geometries, and for two different surface types. We make
use of this feature to refine an earlier estimate of the contact binary
fraction in the Kuiper Belt \citep[KB; see][]{2004AJ....127.3023S}.
\citet{1984A&A...140..265L} considered that lightcurves with ranges between 0.9
and 1.2 magnitudes must be produced by tidally deformed contact binaries
\citep[see also][]{1980Icar...44..807W}. While the maximum range of 1.2$\,$mag is
valid for lunar-type surfaces having negligible limb darkening, our simulations
show that Roche binaries with icy-type surfaces (and thus significant limb
darkening) can produce lightcurve ranges up to 1.57$\,$mag. We searched our
models for the binaries that produce these maximal lightcurve ranges
(1.2$\,$mag for lunar surface and 1.57$\,$mag for icy surface) when observed
equator-on, i.e., at aspect angle $\theta=90\degr$. The aspect angle is
measured between the line of sight and the pole axis of the binary. As the pole
axis of such a binary moves away from the equator-on configuration (as the
aspect angle approaches $0\degr$) the lightcurve range becomes smaller and
smaller; let us denote by $\theta_\mathrm{min}$ the aspect angle at which the
lightcurve range reaches 0.9$\,$mag. If we assume that the pole axes of KB
contact binaries are randomly oriented in space then the detected contact
binary fraction is less than the true fraction by a (geometrical correction)
factor $\cos\theta_\mathrm{min}$. 

\FigGeoCorr

Using our simulations we find the geometrical correction factor to be
$\sim\cos(81.4\degr)=0.15$ for lunar-type surfaces and
$\sim\cos(70.7\degr)=0.33$ for binaries with an icy-type surface (see
Fig.~\ref{Fig.GeoCorr}). We use the fraction of $1/34$ objects with large
($>0.9\,$mag) lightcurve range measured by \citet{2004AJ....127.3023S}, as it
constitutes the largest homogeneous survey for variability. Therefore,
considering only lunar-type surfaces the true fraction of contact binaries is
$f\sim1/(34\times0.15)\sim0.20$. If we consider icy-type surfaces then we
estimate the fraction to be $f\sim1/(34\times0.33)\sim0.09$. Two arguments make
the latter of the two estimates a strong lower limit for the true fraction of
contact (or close) KB binaries. Firstly, the two surface types used here are
simplified limiting models of how real planetary surfaces scatter light. Real
objects presumably exhibit a degree of limb darkening between the two simulated
here.  Secondly, contact binaries with relatively low mass ratios produce
shallower lightcurves which fall below the 0.9$\,$mag threshold adopted here,
and are not accounted for. Our estimate, new in that it includes the effect of
surface scattering, substantiates the idea that a considerable population of
contact/close binary objects in the Kuiper Belt may await discovery
\citep{2004AJ....127.3023S}. 

The Pan-STARRS all-sky survey (http://pan-starrs.ifa.hawaii.edu/) will scan the
entire visible sky, down to $m_R\sim24$, on a weekly basis
\citep{2005AAS...20715004K}. Besides detecting all moving objects to that
brightness limit, this cadence will allow (sparsely sampled) time series
photometric studies, and thus the detection of high variability candidates,
suitable for follow-up observations. The survey will therefore significantly
improve the estimate of the contact binary fraction. The intrinsic fraction of
KB contact binaries can provide important constraints on binary formation
mechanisms \citep[e.g.][]{2002Natur.420..643G} and collisional evolution in the
KB region \citep{2004Icar..168..409P}.

\section{Summary}

Mathematically unique interpretations of rotational lightcurve data are
generally impossible.  Nevertheless, lightcurves can, under physically
plausible assumptions, convey invaluable information about the spins, shapes
and densities of small solar system bodies. In this work we have explored the
role of surface scattering properties on the derivation of bulk densities from
rotational lightcurves, using a quantitative model of rotationally deformed
bodies.  We find that: 

\begin{itemize}

\item With few exceptions, the choice of a particular scattering function does
not strongly affect the densities we obtain from our simulations. Instead, the
presence of surface irregularities (lumps) and some albedo variegation (spots)
on the objects sets the limit to the precision of our density estimates;
surface lumps and spots make it impossible to find {\it one} idealized
equilibrium shape that matches the lightcurve, leading to some degeneracy in
the fits.

\item Our density estimates suggest a trend of increasing density with size.
It is still unclear if such relation is mainly due to composition, to a trend
in porosity, or  to a combination of both.

\end{itemize}

Confirming previous inferences, we find that:

\begin{itemize}

\item The lightcurves of (20000) Varuna and 2003$\,$EL$_{61}$ are well-matched
by rotational equilibrium models in which the bodies are deformed by rotation
into a triaxial shape.  Jacobi ellipsoid models with uniform surface albedo and
a range of limb darkening functions have been used to derive the bulk densities
(Varuna: 992$_{-15}^{+86}\,$kg$\,$m$^{-3}$; 2003$\,$EL$_{61}$:
2551$_{-10}^{+115}\,$kg$\,$m$^{-3}$.

\item The lightcurves of Jovian Trojan (624) Hektor and KBO 2001 QG$_{298}$ are
well-described by contact binary models in which the densities are
2480$_{-80}^{+292}\,$kg$\,$m$^{-3}$ and 590$_{-47}^{+143}\,$kg$\,$m$^{-3}$,
respectively.  

\item The high incidence of KBO lightcurves consistent with a contact binary
interpretation suggests that these bodies are common in the Kuiper belt.  

\end{itemize}

\section*{Acknowledgments}

PL is grateful to the Funda{\c c}\~ao para a Ci\^encia e a Tecnologia
(BPD/SPFH/18828/2004) for financial support. This work was supported, in part,
by a grant from the NSF to DCJ.


\end{document}

%% file: tab1.tex
\begin{deluxetable*}{ccccc}
  \tablecaption{(624) Hektor lightcurve data.
    \label{Table.HektorObs}}
   \tablewidth{0pt}
   \tablehead{
   \colhead{JD\tna} & \colhead{$\theta$\tnb} & \colhead{$\alpha$\tnc} & \colhead{$\Delta m_\mathrm{data}$\tnd} & \colhead{$\Delta m_\mathrm{model}$\tne} \\
   \colhead{} & \colhead{[deg]} & \colhead{[deg]} & \colhead{[mag]} & \colhead{[mag]} 
   }
   \startdata
JD2435989  &	74.9 &	$+4.4$ &       0.775 &	0.737 \\
JD2438795  &	24.8 &	$+4.1$ &       0.113 &	0.063 \\
JD2439556  &	52.5 &	$-5.3$ &       0.398 &	0.302 \\
JD2439977  &	86.3 &	$+4.1$ &       1.055 &	1.048 
   \enddata
  \tablenotetext{a}{Date of observation}
  \tablenotetext{b}{Aspect angle}
  \tablenotetext{c}{Phase angle}
  \tablenotetext{d}{Data lightcurve range}
  \tablenotetext{e}{Best-fit model lightcurve range}
  \tablecomments{Sources cited in Section~\ref{SubSec.Hektor}. }
\end{deluxetable*}

%% file: tab2.tex
\begin{deluxetable*}{ccccccc}
  \tablecaption{List of objects to fit.
    \label{Table.BodyList}}
   \tablewidth{0pt}
   \tablehead{
      \colhead{Object\tna} & \colhead{Family\tnb} & \colhead{$H$ [mag]\tnc}& \colhead{$D_\mathrm{e}$ [km]\tnd} & \colhead{$P$ [hr]\tne} & \colhead{$\Delta m$ [mag]\tnf} & \colhead{$\rho$ [\kgm]\tng} \\
   }
   \startdata
   \cutinhead{\it Triaxial ellipsoids}
	2003$\,$EL$_{61}$  & KBO & 0.2 & 1450 & 3.9154  & $0.28 \pm 0.04$ &  $2585^{+81}_{-44}$ \\
	Varuna             & KBO & 3.2 &  900 & 6.3442  & $0.42 \pm 0.02$ &   $992^{+86}_{-15}$ \\
	2000$\,$GN$_{171}$ & KBO & 6.0 &  360 & 8.329   & $0.61 \pm 0.03$ &   $1946^{+1380}_{-344}$ \\
   \cutinhead{\it Roche binaries}
	2001$\,$QG$_{298}$ & KBO & 6.9 &  240 & 13.7744 & $1.14 \pm 0.04$ &  $590^{+143}_{-47}$ \\
	2000$\,$GN$_{171}$ & KBO & 6.0 &  360 & 8.329   & $0.61 \pm 0.03$ &   $650^{+75}_{-80}$ \\
	(624) Hektor    & Trojan & 7.5 &  180 & 6.9225  & 1.2$\tnh$       & $2480^{+80}_{-292}$ \\
  \enddata
  \tablenotetext{a}{ Object designation.}
  \tablenotetext{b}{ Object family.}
  \tablenotetext{c}{ Absolute magnitude.}
  \tablenotetext{d}{ Approximate equivalent circular diameter.}
  \tablenotetext{e}{ Rotation period.}
  \tablenotetext{f}{ Peak-to-peak lightcurve range.}
  \tablenotetext{g}{ Estimated density.}
  \tablenotetext{h}{ Maximum predicted amplitude, at $\theta=90\deg$.}
  \tablecomments{Sources cited in the text. KBO 2000$\,$GN$_{171}$ is intentionally listed twice, as its nature is uncertain.}
\end{deluxetable*}

%% file: tab3.tex
\begin{deluxetable*}{ccccccc}
  \tablecaption{Roche binary solutions for mass ratio $q=0.25$.
    \label{Table.RocheEll}}
   \tablewidth{0pt}
   \tablehead{
   \colhead{$q$\tna} & \colhead{$B/A$\tnb} & \colhead{$C/A$\tnb} & \colhead{$b/a$\tnc} & \colhead{$c/a$\tnc} & \colhead{$\omega^2/(\pi\,G\,\rho)$\tnd} & \colhead{$l$\tne}
   }
   \startdata
    0.25000 & 0.91674 & 0.83000 & 0.51426 & 0.48000 & 0.10626 & 1.19222 \\
    0.25000 & 0.92292 & 0.84000 & 0.61175 & 0.57000 & 0.10137 & 1.28193 \\
    0.25000 & 0.92891 & 0.85000 & 0.66433 & 0.62000 & 0.09554 & 1.34473 \\
    0.25000 & 0.93473 & 0.86000 & 0.70532 & 0.66000 & 0.08945 & 1.40392 \\
    0.25000 & 0.94037 & 0.87000 & 0.73534 & 0.69000 & 0.08368 & 1.45810 \\
    0.25000 & 0.94584 & 0.88000 & 0.76469 & 0.72000 & 0.07755 & 1.51789 \\
    0.25000 & 0.95114 & 0.89000 & 0.79333 & 0.75000 & 0.07109 & 1.58466 \\
    0.25000 & 0.95629 & 0.90000 & 0.81200 & 0.77000 & 0.06558 & 1.64436 \\
    0.25000 & 0.96128 & 0.91000 & 0.83935 & 0.80000 & 0.05864 & 1.72902 \\
    0.25000 & 0.96612 & 0.92000 & 0.85713 & 0.82000 & 0.05282 & 1.80713 \\
    0.25000 & 0.97081 & 0.93000 & 0.88311 & 0.85000 & 0.04550 & 1.92216 \\
    0.25000 & 0.97537 & 0.94000 & 0.89996 & 0.87000 & 0.03943 & 2.03365 \\
    0.25000 & 0.97979 & 0.95000 & 0.91642 & 0.89000 & 0.03328 & 2.17018 \\
    0.25000 & 0.98407 & 0.96000 & 0.93250 & 0.91000 & 0.02706 & 2.34450 \\
    0.25000 & 0.98824 & 0.97000 & 0.95588 & 0.94000 & 0.01924 & 2.65422 
   \enddata
  \tablenotetext{a}{Mass ratio}
  \tablenotetext{b}{Primary axes ratios}
  \tablenotetext{c}{Secondary axes ratios}
  \tablenotetext{d}{Orbital frequency squared in units of $\pi\,G\,\rho$, where $G$ is the gravitational constant}
  \tablenotetext{e}{Orbital distance in units of $A+a$}
\end{deluxetable*}

%% file: tab4.tex
\begin{deluxetable*}{ccccccccccc}
   \tabletypesize{\tiny}
   \tablecaption{Best three Roche binary model fits to (624) Hektor.
   \label{Table.HektorFit}}
   \tablewidth{0pt}
   \tablehead{
   \colhead{SL\tna} & \colhead{$q$\tnb} & \colhead{$B/A$\tnc}& \colhead{$C/A$\tnc} & \colhead{$b/a$\tnd} & \colhead{$c/a$\tnd} & \colhead{$\omega^2/(\pi\,G\,\rho)$\tne} & \colhead{$d/(A+a)$\tnf} & \colhead{$\rho$\tng} & \colhead{$\chi^2/\chi_\mathrm{best}^2$\tnh}
   }
   \startdata
     Lunar     &0.62&0.80&0.72&0.47&0.43&0.122&0.98& 2480&1.00\\
     Lunar     &0.67&0.77&0.69&0.53&0.49&0.128&1.00& 2374&1.06\\
     Lunar     &0.65&0.79&0.71&0.47&0.43&0.124&0.97& 2453&1.08\\
   \enddata
   \tablenotetext{a}{Surface type.}
   \tablenotetext{b}{Mass ratio of the binary components.}
   \tablenotetext{c}{Axis ratios of primary.}
   \tablenotetext{d}{Axis ratios of the secondary}
   \tablenotetext{e}{Orbital frequency of binary.}
   \tablenotetext{f}{Binary orbital separation.}
   \tablenotetext{g}{Bulk density of the bodies (in \kgm).}
   \tablenotetext{h}{Ratio of $\chi^2$ of model to $\chi^2$ of best-fit model.}
\end{deluxetable*}

%% file: tab5.tex
\begin{deluxetable*}{ccccccccccc}
   \tabletypesize{\tiny}
   \tablecaption{2001$\,$QG$_{298}$ model fit.
   \label{Table.QG}}
   \tablewidth{0pt}
   \tablehead{
   \colhead{S.F.\tna} & \colhead{$\theta$\tnb} & \colhead{$q$\tnc} & \colhead{$B/A$\tnd}& \colhead{$C/A$\tnd} & \colhead{$b/a$\tne} & \colhead{$c/a$\tne} & \colhead{$\omega^2/(\pi\,G\,\rho)$\tnf} & \colhead{$d/(A+a)$\tng} & \colhead{$\rho$\tnh} & \colhead{$\chi^2/\chi_\mathrm{best}^2$\tni}
   }
   \startdata
     \cutinhead{\it Jacobi ellipsoid}
     Lunar     &90&-&0.43&0.34&-&-&0.283&-&  271&2.21\\
     Icy       &90&-&0.56&0.41&-&-&0.327&-&  234&2.50\\
     \\
     Lunar     &75&-&0.43&0.34&-&-&0.283&-&  271&6.60\\
     Icy       &75&-&0.50&0.38&-&-&0.310&-&  248&2.51\\
     \cutinhead{\it Roche binary}
     Lunar     &90&0.84&0.72&0.65&0.45&0.41&0.130&0.90&  590&1.00\\
     Icy       &90&0.44&0.85&0.77&0.53&0.49&0.116&1.09&  659&1.09\\
     \\
     Lunar     &75&1.00&0.44&0.40&0.44&0.40&0.135&0.76&  568&1.62\\
     Icy       &75&0.73&0.74&0.67&0.54&0.49&0.130&0.98&  589&1.16\\
   \enddata
   \tablenotetext{a}{Surface type.}
   \tablenotetext{b}{Aspect angle.}
   \tablenotetext{c}{Mass ratio of the binary components.}
   \tablenotetext{d}{Axis ratios of primary.}
   \tablenotetext{e}{Axis ratios of the secondary}
   \tablenotetext{f}{Spin (orbital) frequency of triaxial ellipsoid (binary).}
   \tablenotetext{g}{Binary orbital separation.}
   \tablenotetext{h}{Bulk density of the bodies (in \kgm).}
   \tablenotetext{i}{Ratio of $\chi^2$ of model to $\chi^2$ of best-fit model.}
\end{deluxetable*}

%% file: tab6.tex
\begin{deluxetable*}{ccccccccccc}
   \tabletypesize{\tiny}
   \tablecaption{2000$\,$GN$_{171}$ model fit.
   \label{Table.GN}}
   \tablewidth{0pt}
   \tablehead{
   \colhead{S.F.\tna} & \colhead{$\theta$\tnb} & \colhead{$q$\tnc} & \colhead{$B/A$\tnd}& \colhead{$C/A$\tnd} & \colhead{$b/a$\tne} & \colhead{$c/a$\tne} & \colhead{$\omega^2/(\pi\,G\,\rho)$\tnf} & \colhead{$d/(A+a)$\tng} & \colhead{$\rho$\tnh} & \colhead{$\chi^2/\chi_\mathrm{best}^2$\tni}
   }
   \startdata
     \cutinhead{\it Jacobi ellipsoid}
     Lunar     &90&-&0.62&0.44&-&-&0.342&-&  613&1.05\\
     Icy       &90&-&0.75&0.50&-&-&0.362&-&  579&1.19\\
     \\
     Lunar     &75&-&0.55&0.41&-&-&0.325&-&  645&1.04\\
     Icy       &75&-&0.71&0.48&-&-&0.357&-&  587&1.20\\
     \cutinhead{\it Roche binary}
     Lunar     &90&0.25&0.92&0.83&0.51&0.48&0.106&1.19& 1972&1.09\\
     Icy       &90&0.25&0.94&0.87&0.74&0.69&0.084&1.46& 2504&1.66\\
     \\
     Lunar     &75&0.34&0.89&0.81&0.45&0.42&0.108&1.09& 1946&1.00\\
     Icy       &75&0.25&0.92&0.84&0.61&0.57&0.101&1.28& 2067&1.05\\
   \enddata
   \tablenotetext{a}{Surface type.}
   \tablenotetext{b}{Aspect angle.}
   \tablenotetext{c}{Mass ratio of the binary components.}
   \tablenotetext{d}{Axis ratios of primary.}
   \tablenotetext{e}{Axis ratios of the secondary}
   \tablenotetext{f}{Spin (orbital) frequency of triaxial ellipsoid (binary).}
   \tablenotetext{g}{Binary orbital separation.}
   \tablenotetext{h}{Bulk density of the bodies (in \kgm).}
   \tablenotetext{i}{Ratio of $\chi^2$ of model to $\chi^2$ of best-fit model.}
\end{deluxetable*}

%% file: tab7.tex
\begin{deluxetable*}{ccccccccccc}
   \tabletypesize{\tiny}
   \tablecaption{2003$\,$EL$_{61}$ model fit.
   \label{Table.EL}}
   \tablewidth{0pt}
   \tablehead{
   \colhead{S.F.\tna} & \colhead{$\theta$\tnb} & \colhead{$q$\tnc} & \colhead{$B/A$\tnd}& \colhead{$C/A$\tnd} & \colhead{$b/a$\tne} & \colhead{$c/a$\tne} & \colhead{$\omega^2/(\pi\,G\,\rho)$\tnf} & \colhead{$d/(A+a)$\tng} & \colhead{$\rho$\tnh} & \colhead{$\chi^2/\chi_\mathrm{best}^2$\tni}
   }
   \startdata
     \cutinhead{\it Jacobi ellipsoid}
     Lunar     &90&-&0.80&0.52&-&-&0.367&-& 2585&1.00\\
     Icy       &90&-&0.88&0.55&-&-&0.372&-& 2551&1.01\\
     \\
     Lunar     &75&-&0.76&0.50&-&-&0.363&-& 2611&1.02\\
     Icy       &75&-&0.86&0.54&-&-&0.371&-& 2557&1.00\\
     \cutinhead{\it Roche binary}
     Lunar     &90&0.25&0.98&0.94&0.90&0.87&0.039&2.03&24049&6.08\\
     Icy       &90&0.25&0.99&0.97&0.96&0.94&0.019&2.65&49286&6.45\\
     \\
     Lunar     &75&0.25&0.95&0.89&0.79&0.75&0.071&1.58&13339&2.56\\
     Icy       &75&0.25&0.97&0.93&0.88&0.85&0.045&1.92&20841&2.46\\
   \enddata
   \tablenotetext{a}{Surface type.}
   \tablenotetext{b}{Aspect angle.}
   \tablenotetext{c}{Mass ratio of the binary components.}
   \tablenotetext{d}{Axis ratios of primary.}
   \tablenotetext{e}{Axis ratios of the secondary}
   \tablenotetext{f}{Spin (orbital) frequency of triaxial ellipsoid (binary).}
   \tablenotetext{g}{Binary orbital separation.}
   \tablenotetext{h}{Bulk density of the bodies (in \kgm).}
   \tablenotetext{i}{Ratio of $\chi^2$ of model to $\chi^2$ of best-fit model.}
\end{deluxetable*}

%% file: tab8.tex
\begin{deluxetable*}{ccccccccccc}
   \tabletypesize{\tiny}
   \tablecaption{(20000) Varuna model fit.
   \label{Table.Va}}
   \tablewidth{0pt}
   \tablehead{
   \colhead{S.F.\tna} & \colhead{$\theta$\tnb} & \colhead{$q$\tnc} & \colhead{$B/A$\tnd}& \colhead{$C/A$\tnd} & \colhead{$b/a$\tne} & \colhead{$c/a$\tne} & \colhead{$\omega^2/(\pi\,G\,\rho)$\tnf} & \colhead{$d/(A+a)$\tng} & \colhead{$\rho$\tnh} & \colhead{$\chi^2/\chi_\mathrm{best}^2$\tni}
   }
   \startdata
     \cutinhead{\it Jacobi ellipsoid}
     Lunar     &90&-&0.69&0.47&-&-&0.354&-& 1020&1.16\\
     Icy       &90&-&0.80&0.52&-&-&0.367&-&  985&1.00\\
     \\
     Lunar     &75&-&0.64&0.45&-&-&0.346&-& 1045&1.17\\
     Icy       &75&-&0.77&0.51&-&-&0.364&-&  992&1.00\\
     \cutinhead{\it Roche binary}
     Lunar     &90&0.25&0.92&0.84&0.61&0.57&0.101&1.28& 3563&5.48\\
     Icy       &90&0.25&0.95&0.88&0.76&0.72&0.078&1.52& 4657&8.12\\
     \\
     Lunar     &75&0.25&0.92&0.83&0.51&0.48&0.106&1.19& 3399&2.95\\
     Icy       &75&0.25&0.93&0.85&0.66&0.62&0.096&1.34& 3780&3.53\\
   \enddata
   \tablenotetext{a}{Surface type.}
   \tablenotetext{b}{Aspect angle.}
   \tablenotetext{c}{Mass ratio of the binary components.}
   \tablenotetext{d}{Axis ratios of primary.}
   \tablenotetext{e}{Axis ratios of the secondary}
   \tablenotetext{f}{Spin (orbital) frequency of triaxial ellipsoid (binary).}
   \tablenotetext{g}{Binary orbital separation.}
   \tablenotetext{h}{Bulk density of the bodies (in \kgm).}
   \tablenotetext{i}{Ratio of $\chi^2$ of model to $\chi^2$ of best-fit model.}
\end{deluxetable*}

%% file: msxxx061130.bbl
\begin{thebibliography}{41}
\expandafter\ifx\csname natexlab\endcsname\relax\def\natexlab#1{#1}\fi

\bibitem[{{Belskaya} {et~al.}(2006){Belskaya}, {Ortiz}, {Rousselot}, {Ivanova},
  {Borisov}, {Shevchenko}, \& {Peixinho}}]{2006Icar..184..277B}
{Belskaya}, I.~N., {Ortiz}, J.~L., {Rousselot}, P., {Ivanova}, V., {Borisov},
  G., {Shevchenko}, V.~G., \& {Peixinho}, N. 2006, Icarus, 184, 277

\bibitem[{{Cellino} {et~al.}(1989){Cellino}, {Zappala}, \&
  {Farinella}}]{1989Icar...78..298C}
{Cellino}, A., {Zappala}, V., \& {Farinella}, P. 1989, Icarus, 78, 298

\bibitem[{{Chandrasekhar}(1963)}]{1963ApJ...138.1182C}
{Chandrasekhar}, S. 1963, \apj, 138, 1182

\bibitem[{{Chandrasekhar}(1969)}]{1969efe..book.....C}
---. 1969, {Ellipsoidal figures of equilibrium} (The Silliman Foundation
  Lectures, New Haven: Yale University Press, 1969)

\bibitem[{{Chandrasekhar} \& {Lebovitz}(1962)}]{1962ApJ...136.1037C}
{Chandrasekhar}, S. \& {Lebovitz}, N.~R. 1962, \apj, 136, 1037

\bibitem[{{Cruikshank}(1977)}]{1977Icar...30..224C}
{Cruikshank}, D.~P. 1977, Icarus, 30, 224

\bibitem[{{Cruikshank}(2005)}]{2005AdSpR..36.1070C}
---. 2005, Advances in Space Research, 36, 1070

\bibitem[{{de Angelis}(1995)}]{1995P&SS...43..649D}
{de Angelis}, G. 1995, \planss, 43, 649

\bibitem[{{Degewij} {et~al.}(1979){Degewij}, {Tedesco}, \&
  {Zellner}}]{1979Icar...40..364D}
{Degewij}, J., {Tedesco}, E.~F., \& {Zellner}, B. 1979, Icarus, 40, 364

\bibitem[{{Dunlap} \& {Gehrels}(1969)}]{1969AJ.....74..796D}
{Dunlap}, J.~L. \& {Gehrels}, T. 1969, \aj, 74, 796

\bibitem[{{Fern{\'a}ndez} {et~al.}(2003){Fern{\'a}ndez}, {Sheppard}, \&
  {Jewitt}}]{2003AJ....126.1563F}
{Fern{\'a}ndez}, Y.~R., {Sheppard}, S.~S., \& {Jewitt}, D.~C. 2003, \aj, 126,
  1563

\bibitem[{{Goldreich} {et~al.}(2002){Goldreich}, {Lithwick}, \&
  {Sari}}]{2002Natur.420..643G}
{Goldreich}, P., {Lithwick}, Y., \& {Sari}, R. 2002, \nat, 420, 643

\bibitem[{{Hartmann} \& {Cruikshank}(1978)}]{1978Icar...36..353H}
{Hartmann}, W.~K. \& {Cruikshank}, D.~P. 1978, Icarus, 36, 353

\bibitem[{{Hartmann} \& {Cruikshank}(1980)}]{1980Sci...207..976H}
---. 1980, Science, 207, 976

\bibitem[{{Hicks} {et~al.}(2005){Hicks}, {Simonelli}, \&
  {Buratti}}]{2005Icar..176..492H}
{Hicks}, M.~D., {Simonelli}, D.~P., \& {Buratti}, B.~J. 2005, Icarus, 176, 492

\bibitem[{{Jeans}(1919)}]{1919QB981.J4.......}
{Jeans}, J.~H. 1919, {Problems of cosmogony and stellar dynamics} (Cambridge,
  University press, 1919.)

\bibitem[{{Jewitt}(2002)}]{2002acm..conf...11J}
{Jewitt}, D. 2002, in ESA SP-500: Asteroids, Comets, and Meteors: ACM 2002, ed.
  B.~{Warmbein}, 11--19

\bibitem[{{Jewitt}(in print)}]{2006saas.conf.....J}
{Jewitt}, D. in print, in Saas-Fee Advanced Course 35: Trans-Neptunian Objects
  and Comets, ed. D.~{Jewitt}, A.~{Morbidelli}, \& H.~{Rauer} (Springer)

\bibitem[{{Jewitt} {et~al.}(2001){Jewitt}, {Aussel}, \&
  {Evans}}]{2001Natur.411..446J}
{Jewitt}, D., {Aussel}, H., \& {Evans}, A. 2001, \nat, 411, 446

\bibitem[{{Jewitt} \& {Luu}(1990)}]{1990AJ....100..933J}
{Jewitt}, D.~C. \& {Luu}, J.~X. 1990, \aj, 100, 933

\bibitem[{{Jewitt} \& {Sheppard}(2002)}]{2002AJ....123.2110J}
{Jewitt}, D.~C. \& {Sheppard}, S.~S. 2002, \aj, 123, 2110

\bibitem[{{Johnson} \& {Lunine}(2005)}]{2005Natur.435...69J}
{Johnson}, T.~V. \& {Lunine}, J.~I. 2005, \nat, 435, 69

\bibitem[{{Kaasalainen} {et~al.}(2002){Kaasalainen}, {Mottola}, \&
  {Fulchignoni}}]{2002aste.conf..139K}
{Kaasalainen}, M., {Mottola}, S., \& {Fulchignoni}, M. 2002, Asteroids III, 139

\bibitem[{{Kaasalainen} \& {Torppa}(2001)}]{2001Icar..153...24K}
{Kaasalainen}, M. \& {Torppa}, J. 2001, Icarus, 153, 24

\bibitem[{{Kaasalainen} {et~al.}(2001){Kaasalainen}, {Torppa}, \&
  {Muinonen}}]{2001Icar..153...37K}
{Kaasalainen}, M., {Torppa}, J., \& {Muinonen}, K. 2001, Icarus, 153, 37

\bibitem[{{Kaiser} \& {Pan-STARRS Team}(2005)}]{2005AAS...20715004K}
{Kaiser}, N. \& {Pan-STARRS Team}. 2005, Bulletin of the American Astronomical
  Society, 37, 1409

\bibitem[{{Leone} {et~al.}(1984){Leone}, {Paolicchi}, {Farinella}, \&
  {Zappala}}]{1984A&A...140..265L}
{Leone}, G., {Paolicchi}, P., {Farinella}, P., \& {Zappala}, V. 1984, \aap,
  140, 265

\bibitem[{{Li}(2005)}]{2005PhDT.........7L}
{Li}, J.-Y. 2005, Ph.D.~Thesis

\bibitem[{{Marchis} {et~al.}(2006{\natexlab{a}}){Marchis}, {Hestroffer},
  {Descamps}, {Berthier}, {Bouchez}, {Campbell}, {Chin}, {van Dam}, {Hartman},
  {Johansson}, {Lafon}, {Le Mignant}, {de Pater}, {Stomski}, {Summers},
  {Vachier}, {Wizinovich}, \& {Wong}}]{2006Natur.439..565M}
{Marchis}, F., {Hestroffer}, D., {Descamps}, P., {Berthier}, J., {Bouchez},
  A.~H., {Campbell}, R.~D., {Chin}, J.~C.~Y., {van Dam}, M.~A., {Hartman},
  S.~K., {Johansson}, E.~M., {Lafon}, R.~E., {Le Mignant}, D., {de Pater}, I.,
  {Stomski}, P.~J., {Summers}, D.~M., {Vachier}, F., {Wizinovich}, P.~L., \&
  {Wong}, M.~H. 2006{\natexlab{a}}, \nat, 439, 565

\bibitem[{{Marchis} {et~al.}(2006{\natexlab{b}}){Marchis}, {Wong}, {Berthier},
  {Descamps}, {Hestroffer}, {Vachier}, {Le Mignant}, {Keck Observatory}, \& {de
  Pater}}]{2006IAUC.8732....1M}
{Marchis}, F., {Wong}, M.~H., {Berthier}, J., {Descamps}, P., {Hestroffer}, D.,
  {Vachier}, F., {Le Mignant}, D., {Keck Observatory}, W.~M., \& {de Pater}, I.
  2006{\natexlab{b}}, \iaucirc, 8732, 1

\bibitem[{{Millis}(1977)}]{1977Icar...31...81M}
{Millis}, R.~L. 1977, Icarus, 31, 81

\bibitem[{{Person} {et~al.}(2006){Person}, {Elliot}, {Gulbis}, {Pasachoff},
  {Babcock}, {Souza}, \& {Gangestad}}]{2006AJ....132.1575P}
{Person}, M.~J., {Elliot}, J.~L., {Gulbis}, A.~A.~S., {Pasachoff}, J.~M.,
  {Babcock}, B.~A., {Souza}, S.~P., \& {Gangestad}, J.~W. 2006, \aj, 132, 1575

\bibitem[{{Petit} \& {Mousis}(2004)}]{2004Icar..168..409P}
{Petit}, J.-M. \& {Mousis}, O. 2004, Icarus, 168, 409

\bibitem[{{Rabinowitz} {et~al.}(2006){Rabinowitz}, {Barkume}, {Brown}, {Roe},
  {Schwartz}, {Tourtellotte}, \& {Trujillo}}]{2006ApJ...639.1238R}
{Rabinowitz}, D.~L., {Barkume}, K., {Brown}, M.~E., {Roe}, H., {Schwartz}, M.,
  {Tourtellotte}, S., \& {Trujillo}, C. 2006, \apj, 639, 1238

\bibitem[{{Russell}(1906)}]{1906ApJ....24....1R}
{Russell}, H.~N. 1906, \apj, 24, 1

\bibitem[{{Sawyer}(1991)}]{1991PhDT.......105S}
{Sawyer}, S.~R. 1991, Ph.D.~Thesis

\bibitem[{{Sheppard} \& {Jewitt}(2004)}]{2004AJ....127.3023S}
{Sheppard}, S.~S. \& {Jewitt}, D. 2004, \aj, 127, 3023

\bibitem[{{Sheppard} \& {Jewitt}(2002)}]{2002AJ....124.1757S}
{Sheppard}, S.~S. \& {Jewitt}, D.~C. 2002, \aj, 124, 1757

\bibitem[{{Takahashi} \& {Ip}(2004)}]{2004PASJ...56.1099T}
{Takahashi}, S. \& {Ip}, W.-H. 2004, \pasj, 56, 1099

\bibitem[{{Weidenschilling}(1980)}]{1980Icar...44..807W}
{Weidenschilling}, S.~J. 1980, Icarus, 44, 807

\bibitem[{{Wijesinghe} \& {Tedesco}(1979)}]{1979Icar...40..383W}
{Wijesinghe}, M.~P. \& {Tedesco}, E.~F. 1979, Icarus, 40, 383

\end{thebibliography}
